%% file: sample-sigconf.tex
\documentclass[sigconf,anonymous=false]{acmart}
\AtBeginDocument{%
  \providecommand\BibTeX{{%
    \normalfont B\kern-0.5em{\scshape i\kern-0.25em b}\kern-0.8em\TeX}}}

\setcopyright{acmcopyright}
\copyrightyear{2021}
\acmYear{2021}

\acmConference[Woodstock '18]{Woodstock '18: ACM Symposium on Neural
  Gaze Detection}{June 03--05, 2018}{Woodstock, NY}
\acmBooktitle{Woodstock '18: ACM Symposium on Neural Gaze Detection,
  June 03--05, 2018, Woodstock, NY}
\acmPrice{15.00}
\acmISBN{978-1-4503-XXXX-X/18/06}

\usepackage{tikz}
\usetikzlibrary{bayesnet}
\usepackage{booktabs}
\usepackage{array}
\usepackage{subcaption}
\usepackage{bm}
\usepackage{algorithm}
\usepackage{algorithmic}
\newtheorem{theorem}{Theorem}

\begin{document}

\title{Identifying Coordinated Accounts on Social Media through Hidden Influence and Group Behaviours}

\author{Karishma Sharma}
\authornote{These authors contributed equally. They are also the corresponding authors.}
\affiliation{University of Southern California \country{USA}}
\email{krsharma@usc.edu}

\author{Yizhou Zhang}
\authornotemark[1]
\affiliation{University of Southern California \country{USA}}
\email{zhangyiz@usc.edu}

\author{Emilio Ferrara}
\affiliation{University of Southern California \country{USA}}
\email{emiliofe@usc.edu}

\author{Yan Liu}
\affiliation{University of Southern California \country{USA}}
\email{yanliu.cs@usc.edu}

\begin{abstract}
  Disinformation campaigns on social media, involving coordinated activities from malicious accounts towards manipulating public opinion, have become increasingly prevalent. Existing approaches to detect coordinated accounts either make very strict assumptions about coordinated behaviours, or require part of the malicious accounts in the coordinated group to be revealed in order to detect the rest. To address these drawbacks, we propose a generative model, AMDN-HAGE (Attentive Mixture Density Network with Hidden Account Group Estimation) which jointly models account activities and hidden group behaviours based on Temporal Point Processes (TPP) and Gaussian Mixture Model (GMM), to capture inherent characteristics of coordination which is, accounts that coordinate must strongly influence each other’s activities, and collectively appear anomalous from normal accounts. 
  To address the challenges of optimizing the proposed model, we provide a bilevel optimization algorithm with theoretical guarantee on convergence. 
  We verified the effectiveness of the proposed method and training algorithm on real-world social network data collected from Twitter related to coordinated campaigns from Russia's Internet Research Agency targeting the 2016 U.S. Presidential Elections, and to identify coordinated campaigns related to the COVID-19 pandemic.  
  Leveraging the learned model, we find that the average influence between coordinated account pairs is the highest.
  On COVID-19, we found coordinated group spreading anti-vaccination, anti-masks conspiracies that suggest the pandemic is a hoax and political scam. 
\end{abstract}





\maketitle

\input{introduction}

\input{related}
\input{method}

\input{expts}

\input{expts_covid}

\input{conclusion}

\begin{acks}
This work is supported by NSF Research Grant (IIS-1254206). Views and conclusions are of the authors and should not be interpreted as representing the social policies of the funding agency, or U.S. Government. Emilio Ferrara acknowledges support by the Air Force Office for Scientific Research (grant no. FA9550-20-1-0224).
\end{acks}

\bibliographystyle{ACM-Reference-Format}
{\footnotesize\bibliography{main}}

\appendix
\input{suppl}

\end{document}

%% file: introduction.tex
\section{Introduction}

\begin{figure}
    \centering
    \includegraphics[width=0.5\textwidth,height=6cm]{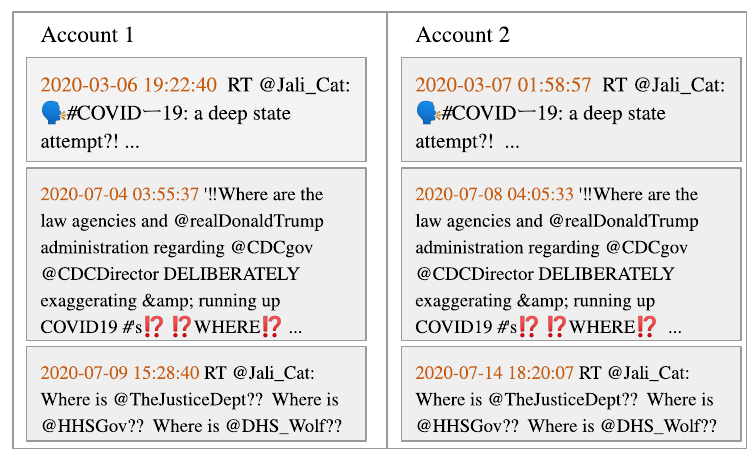}
    \caption{Coordinated accounts suspended by Twitter in COVID-19 data (also identified with proposed method). Example tweets from these accounts show political conspiracies, promoted in coordination. The time difference of their coordinated activity varies from less than 6 hours in one instance to more than half a week in the other.}
    \label{fig:motivating_example}
\end{figure}


In recent times, the persistent abuse of social media for spreading disinformation and influencing public opinion and social outcomes has become an increasingly pressing problem \cite{sharma2019combating}. It has largely been used as a tactic to influence elections \cite{badawy2019characterizing}, public perception on social issues such as social distancing policies related to COVID-19 \cite{sharma2020coronavirus} and other local and global events. 
The issue has gained even more relevance during the ongoing COVID-19 pandemic, where increased reliance on social media for information related to healthcare and policies, has made it an easy target for large-scale disinformation campaigns \cite{sharma2020coronavirus}. The earliest reported cases of disinformation campaigns surfaced when investigations by the U.S. congress found coordinated accounts operated by Russia's “Internet Research
Agency” (IRA) or Russian ``troll farm" on social media to influence the 2016 U.S. Election by promotion of disinformation, and politically divisive narratives \cite{badawy2019characterizing}. Identification of coordinated accounts used to manipulate social media is critical.



To address this crucial task, earlier approaches have tried to uncover coordinated accounts based on their \emph{individual} behaviours, from account features and participation in disinformation promotion \cite{addawood2019linguistic}, or collective \emph{group} behaviours, such as activities synchronized in time \cite{cao2014uncovering,pacheco2020uncovering}. 
However, they all face inherent limitations.
The methods relying on disinformation cues or automated behaviours to detect coordinated actors fall short in detecting human operated accounts that instead use persuasion, defamation, and polarization to manipulate opinions (characteristics noted in coordinated influence operations \cite{martin2019recent}). 

Existing methods exploiting collective behaviours heavily rely on assumed features of coordinated behaviour e.g. similar sequence of hashtags as coordination signature or activities synchronized in time \cite{pacheco2020uncovering,yu2015glad,wang2016unsupervised}. However such features can be inconsistent \cite{zannettou2019let}, limiting generalization to unseen accounts. Moreover, reliance on hand-crafted coordination signatures can only capture a limited range of behaviours, and are ineffective at reducing false positives, making strict assumptions on coordinated groups which need not hold true. Fig. \ref{fig:motivating_example} shows an example of accounts officially suspended by Twitter and their activities observed in COVID-19 data (also identified as coordinated with our method).
The time differences of coordinated activities observed from the accounts is diverse, with less than 6 hours in one case to more than half a week in another. 

In this work, to address these shortcomings we propose to model the following inherent characteristics of coordination:
\begin{itemize}
    \item Strong hidden influence. If accounts coordinate to amplify social media posts or target specific individuals, there should be a strong hidden (latent) influence between their activities. Significant recent computational propaganda is produced and operated by political operatives and governments \cite{woolley2018computational}.
    \item Highly concerted activities. The collective behaviours of coordinated accounts should be collectively anomalous, from other normal accounts on the network with less organized activity patterns (i.e., observations that deviate from normal when generated by a different mechanism \cite{hawkins1980identification}).
\end{itemize}
To capture them,
we propose \textbf{AMDN-HAGE (Attentive Mixture Density Network with Hidden Account Group Estimation)}, an unsupervised generative model for identification of coordinated accounts, which jointly models account activities
and hidden account groups based on Neural Temporal Point
Processes (NTPP) and Gaussian Mixture Model (GMM). To learn the latent interactions or influence between account activities, we model the distribution of future activities conditioned on past activities of all accounts with temporal differences, and jointly capture collective anomalous behaviour by simultaneously learning the group membership of accounts.
To address the joint optimization, we propose a bilevel optimization algorithm using both stochastic gradient descent and expectation-maximization, from observed activity traces. 

As AMDN-HAGE directly learns account representations and hidden groups from activity traces with only account ids and activity timestamps in an unsupervised manner, it does not require knowledge of partially uncovered accounts in coordinated groups, or predefined individual features, although they can also be plugged in easily if necessary.
Furthermore, unlike the existing models relying on strict group behaviour hypothesises like timestamp synchronization, our model only assumes that coordinated accounts are an anomalous group with concerted activities, allowing more data-driven identification of coordinated groups.
In addition to above advantages on effectiveness, by incorporating an explicit attention module, we enable our model to learn and output the strength of latent interactions between accounts on the network. 
In general, our contributions are the following,
\begin{itemize}
    \item We propose AMDN-HAGE to detect coordinated campaigns from collective group behaviours inferred from account activities based on NTTP and GMM. 
    \item We provide a bilevel optimization algorithm for joint learning
    of activity trace modeling (NTPP) and social group modeling (GMM) 
    with theoretical and empirical guarantee. 
    \item Extensive experiments on real-world Twitter data about known coordinated campaign 
    verify our method is highly effective on coordination detection. From learned model we find identifiable patterns of coordinated accounts such as the influence between them is highest but also decay fastest.
    \item We apply the method to identify unknown coordinated campaigns in COVID-19 data, and find coordinated group of Spanish and English accounts (NoMask, NoVaccine, NoALaVacuna, NoAlNuevoOrdenMundia, QAnon) about no masks, no vaccine, no new world order conspiracies, opposing Bill Gates, that suggest COVID-19 is a hoax and political scam.
\end{itemize}

%% file: related.tex
\section{Related Work}

The problem of disinformation and social media abuse has reduced trust in online platforms. Efforts to combat false and misleading information have been widely studied in many different contexts \cite{sharma2019combating} ranging from detection of false information from content features and the responses to it on social media, to understanding the diffusion patterns \cite{icwsmNetInference} and accounts involved in its spread \cite{ferrara2016rise}. 
Different from social bots and individual malicious accounts, a growing area of research is the detection of coordinated accounts (coordinated disinformation or influence campaigns), also called ``troll farms", orchestrated by group of human and/or bot accounts that are made to work jointly to promote disinformation or other narratives \cite{luceri2020detecting}. The main techniques to identify coordinated accounts are:

\textbf{Individual behaviours.} Existing works mainly uses two kinds of individual behaviours. The first one is the participation in disinformation spreading \cite{icwsmNetInference,ruchansky2017csi}. For instance, \citeauthor{ruchansky2017csi} propose a fake news detection model that assigns a suspiciousness score to accounts based on their participation in fake news cascades \cite{ruchansky2017csi}. The second kind are individual characteristics such as deceptive linguistic features, number of shared links, hashtags and device of posting and cross-platform activity \cite{addawood2019linguistic,im2020still,zannettou2019disinformation}. Apart from above pre-defined features, the activity traces of troll accounts have been found useful for understanding malicious behaviours. In recent work, the tweet, retweet and reply patterns of Twitter accounts are utilized to infer the incentives or rewards behind their activities, formulated as an inverse reinforcement learning problem \cite{luceri2020detecting}. Based on the estimated rewards, the authors found that the behaviours of trolls was different from regular users as they appeared to perform their activity regardless of the responses.

\textbf{Collective behaviours.} Approaches that examine the collective or group behaviours as a whole to detect anomalous malicious accounts are related to our approach. \citeauthor{cao2014uncovering} and \citeauthor{gupta2019malreg} cluster accounts that take similar actions around the same time, based on the assumption that malicious account activities are synchronized in time \cite{cao2014uncovering,gupta2019malreg}. Other works cluster or partition a account similarity graph defined over hand-crafted features assumed to be indicative of coordinated behaviours, including the sequence of hashtags or articles shared collectively by a large group of accounts \cite{pacheco2020uncovering,wang2016unsupervised}. The significant limitation of such approaches is that the assumption on synchronization or hand-crafted features used to define coordination might not hold. In contrast, we propose to automatically learn and detect coordinated behaviours from observed account activities by learning the latent account interactions.

%% file: method.tex
\section{Task Definition and Preliminaries}

\subsection{Task Definition}
In this section, we introduce the task of detecting coordinated accounts in social networks from collective group behaviours of the accounts. Coordinated campaigns are orchestrated efforts where accounts collude to inorganically spread and amplify the spread of specific narratives for opinion manipulation, and the task we address is to identify such coordinated accounts. 

In this work, we propose AMDN-HAGE (Attentive Mixture Density Network with Hidden Account Group Estimation), an unsupervised generative model for coordination detection. It jointly models account \textbf{activity traces} and latent \textbf{account groups}, to learn collective group behaviours from observed account activities, and detect coordinated accounts with collective anomalous behaviours.

\textbf{Activity traces:} The only input we consider are the activity traces of accounts on the social network. An activity trace can be represented as a sequence of events ordered in time, which can be formulated as $C_s = [(u_1, t_1), (u_2, t_2), (u_3, t_3), \cdots (u_n, t_n)]$. Each tuple $(u_i, t_i)$ corresponds to an activity by account $u_i$ at time $t_i$. The activities represent account actions on the network such as posting original content, re-sharing, replying, or reacting to other posts. 

In order to provide platform/language independent detection, we do not include the type of action, or features such as content of the post, and account metadata, although additional available features can be easily incorporated in the method. The basic input is the most easily available for any social network. Furthermore, the only assumption we make on the coordinated campaign is that:
\begin{itemize}
    \item Compared to normal accounts, the number of coordinated accounts is quite small (i.e., collectively anomalous).
    \item Coordinated users have highly concerted activity patterns
\end{itemize}

\textbf{Hidden Account Group:} In real social networks, accounts with similar activities form social groups, that can constitute normal communities as well as coordinated groups. Supposing that there are $N$ groups in the account set $U$, we can define a membership function $M:U\rightarrow \{1,\cdots,N\}$, which projects each account $u_i$ to its group index. This membership in many cases is hidden or unknown \cite{baumes2004discovering}. Acquiring $M$ can help us identify collective anomalous group behaviours to detect coordinated groups. In this work, we aim to learn the hidden groups from only the observed activity traces.


\subsection{Temporal Point Process}

A temporal point process (TPP) is a stochastic process
whose realization is a sequence of discrete events in continuous time $t \in \mathbb{R}^{+}$ \cite{daley2007introduction}. 
The history of events in the sequence up to time $t$ are generally denoted as $H_t = \{(u_i, t_i) | t_i < t, u_i \in \mathcal{U} \}$ where $\mathcal{U}$ represents the set of event types (here, accounts). The conditional intensity function $\lambda(t|H_t)$ of a point process is defined as the instantaneous rate of an event in an infinitesimal window at time $t$ given the history i.e. $\lambda(t|H_t)dt= \mathbb{E}[dN(t)|H_t]$ where $N(t)$ is the number of events up to time $t$. 
The conditional density function of the $i^{\textrm{th}}$ event can be derived from the conditional intensity \cite{du2016recurrent} as 
\begin{equation}
    p(t|H_t) = \lambda(t|H_t) \exp \left( -\int_{t_{i-1}}^{t} \lambda (s|H_t) ds \right)
\end{equation}
In social network data, the widely used formulation of the conditional intensity is the multivariate Hawkes Process (HP) \cite{zhou2013learning}, defined as $\lambda_i(t|H_t) = \mu_i +  \sum_{t_j < t} \alpha_{i,j} \kappa(t - t_j)$, 
where $\lambda_i(t|H_t)$ is the conditional intensity of event type $i$ at time $t$ with base intensity $\mu_i > 0$ and mutually triggering intensity $\alpha_{i,j} > 0$ capturing the \textbf{influence} of event type $j$ on $i$ and $\kappa$ is a decay kernel to model \emph{influence decay} over time. $\mu$ and $\alpha$ are learnable parameters. Since HP's fixed formulation and few learnable parameters limit its expressive power, recent works propose to model the intensity function with neural networks \cite{du2016recurrent,mei2017neural,zhangself,zuo2020transformer,shchur2019intensity,omi2019fully}. 

\section{Coordination Detection Method}

In order to capture the latent influence between account's activities, and collective behaviours of coordinated groups, as well as diversity in coordinated activities from such accounts, we introduce the proposed model AMDN-HAGE. 

\begin{figure}
    \centering
    \includegraphics[width=3.3in]{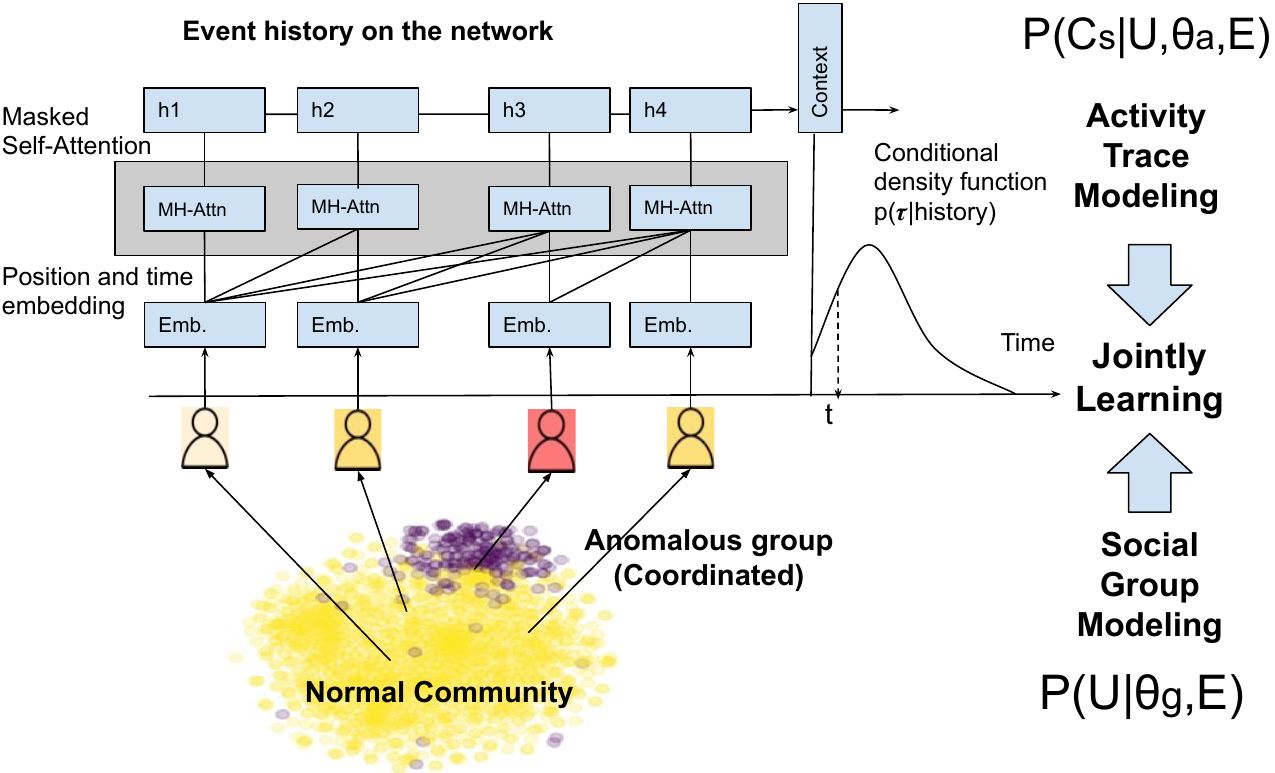}
    \caption{Architecture of proposed (AMDN-HAGE) to model conditional density function of account activities and hidden groups on social media.}
    \label{fig:model_arch}
\end{figure}

AMDN-HAGE consists of two components: an Attentive Mixture Density Network (AMDN) that models observed activity traces as a temporal point process and a Hidden Account Group Estimation (HAGE) component that models account groups as mixture of multiple distributions. An overview is shown in Figure \ref{fig:model_arch}. The two components share the account embedding layer and reflect the complete generative process that the accounts are first drawn from multiple hidden groups and then interact with each other so that activity traces are observed. Using the observed activity traces, we can learn the generative model by maximizing the likelihood function of the joint model, and acquire not only account embeddings but also a activity trace model and group membership function. Denoting the account embeddings as $E$, the parameters in AMDN as $\theta_a$ and the parameters in HAGE as $\theta_g$, the joint likelihood function can be written as:
\begin{equation}
\begin{aligned}
    \log p(C_s, U;\theta_g,\theta_a,E)& = \log p(C_s| U;\theta_g,\theta_a,E) + \log p(U;\theta_g,\theta,E) \\
    &= \log p(C_s| U;\theta_a,E) + \log p(U;\theta_g,E)\\
\end{aligned}
\end{equation}

$p(C_s| U;\theta_a,E)$ is the probability density that the activity traces are observed given a known account set, and $p(U;\theta_g,E)$ is the probability density that we observe the account set drawn from the latent hidden social groups. With the account embeddings and the learned membership, we obtain collectively anomalous groups as latent groups with anomalous distributions having small variance or size compared to the rest of the accounts, to detect coordination.


\subsection{Modeling Account Activities} 
In this section, we introduce the AMDN component (Attentive Mixture Density Network) to model account activities for coordination detection. AMDN consists of two parts: a history encoder and an event decoder. Suppose we are modeling the activity $(u_i,t_i)$, the history encoder represents all the activities that happened before $t_i$ as a vector $H_{t_i}$. Then the event decoder, which is the conditional density function, predicts $(u_i,t_i)$ based on the history representation $H_{t_i}$ and the known account set. This encoder-decoder architecture models activities with likelihood $p(C_s| U;\theta_a,E)$ factorized as:
\begin{equation}
    \log p(C_s|U;\theta_a,E)
    = \sum_{i=1}^{L} \left[ \log p_{\theta_a,E}(t_i|H_{t_i}) + \log p_{\theta_a,E}(u_i|H_{t_i}) \right]
    \label{eq:amdn_likelihood}
\end{equation}
We provide architecture details in the following paragraphs. 


\subsubsection{AMDN architecture and training}

\begin{table}[t]
    \centering
    \renewcommand*{\arraystretch}{0.8}
    \caption{Summary of neural point process models}
    \label{tab:tpp}
    \begin{tabular}{l|p{1.2cm}|p{1cm}|p{1cm}|p{1cm}}
    \toprule
    Model & Flexible intensity function & 
    Closed-form likelihood &
    Long-range dependencies &  Inter-pretable Influence  \\
    \midrule
    HP \cite{zhou2013learning} & no & y & y & y \\
    RMTPP \cite{du2016recurrent} & limited & y & limited & no \\
    FullyNN \cite{omi2019fully} & y & y & limited & no \\
    LogNormMix \cite{shchur2019intensity} & y & y & limited & no \\
    SAHP \cite{zhangself} & limited & no & y & y \\
    THP \cite{zuo2020transformer} & y & no & y & y \\
    \midrule
    \textbf{AMDN} & y & y & y & y \\ 
    \bottomrule
    \end{tabular}
\end{table}

In Table~\ref{tab:tpp}, we summarize existing point process models (detailed in Appx.); these models suffer from different drawbacks with a trade-off on flexible intensity function (better expressive power), closed-form likelihood (reducing gradient noise in training) and  interpretable influence (explicit influence score on event pairs). For modeling coordinated accounts, above properties are all useful. Thus, we develop AMDN, which has all the above properties. We use masked self-attention \cite{vaswani2017attention} (with additional temporal encoding for handling irregular inter-event times) to encode the event history for interpretable influence between past and future events (alternative to the recurrent neural network used in \cite{shchur2019intensity}), but still use a log-normal mixture distribution as event decoder to model the conditional density of the next event given the history (similar to \cite{shchur2019intensity}), achieving all properties.


\textbf{Encoding event sequences with masked self-attention.} Let $\tau \in \mathbb{R}^+$ represent inter-event time, $p(\tau|H_\tau)$ the conditional density. History $H_\tau$ is encoded with a neural network to automatically extract useful features, similar to other neural point process models. 

\paragraph{Masked self-attention with position encoding} For interpretable influence of past event on future events, we encode the event sequence with masked self-attention \cite{vaswani2017attention}. 
\begin{align}
\begin{split}
& A = \sigma (QK^T/\sqrt{d})
   \quad\mathrm{and}\quad 
H_{attn} = AV
\quad \\
& Q = X W_q, \hspace*{0.1cm} K = X W_k, \hspace*{0.1cm} V = X W_v
\end{split}
\end{align}
with masked attention weights $A$ of pairwise influence from previous events, 
input sequence representation $X \in \mathbb{R}^{L \times d}$ ($L$ sequence length, $d$ feature dimension), and learnable weights $W_q, W_k, W_v$. 
At the end, we apply layer normalization, dropout and feed-forward layer to $H_{attn}$ to get output $H_{out} \in \mathbb{R}^{L \times d}$. To maintain ordering of the history events, we represent the position information of the $i$-th event as an $m$-dim position encoding $PE_{pos=i}$ with trigonometric integral function following \cite{vaswani2017attention}.



\paragraph{Account and time encoding} 

We also represent account type and time $(u_i, t_i)$ using
learnable account embedding matrix $E \in \mathbb{R}^{|U| x m}$ and translation-invariant temporal
kernel functions \cite{xu2019self}, 
using feature maps $\phi$ with multiple periodic functions of different frequencies $\omega$ to embed inter-event times.
\begin{equation}
    \phi_{\omega}(t) = [\sqrt{c_1}, \cdots \sqrt{c_{2j}} \cos(j\pi t/\omega), \sqrt{c_{2j+1}} \sin(j\pi t/\omega) \cdots]
\end{equation}
\begin{equation}
     \phi(t) = [\phi_{\omega_1}(t), \phi_{\omega_2}(t) \cdots \phi_{\omega_k}(t)]^T
 \end{equation}

The input $X_i$ (to the attention head) of the $i^{th}$ event $(u_i, t_i)$, is a concatenation of event, position and temporal embedding,
\begin{equation}
    X_i = [E_{u_i}, PE_{pos=i}, \phi(t_i - t_{i-1})]
\end{equation}

\paragraph{Event history context vector} The attention mechanism gives us representations of each event using attention over events prior to it. We can use the representation of the last event or a recurrent network layer over the attention outputs $H_{out} \in \mathbb{R}^{Lxd}$ to summarize events histories into context vectors $C \in \mathbb{R}^{Lxd}$ where $L$ is event sequence length. Each $c_i \in C$ is a context vector encoding history of events up to $t_i$ i.e. history $H_{t_i}$ of the temporal point process.
\\\\
\textbf{Conditional probability density function.}  With the encoded event history (context vector), the event decoder (learnable conditional density function $p(\tau|H_\tau)$) is used to generate the distribution of the next event time conditioned on the history.

While we can choose any functional form for $p(\tau|H_\tau)$), the only condition is that it should be a valid PDF (non-negative, and integrate to $1$ over $\tau \in \mathbb{R}_+$). To maintain a valid PDF, exponential or other distributions with learnable parameters are generally used in point process models \cite{zhou2013learning,du2016recurrent,shchur2019intensity}. We define the PDF as mixture of log-normal distributions since the domain of $\tau \in \mathbb{R}_{+}$ is non-negative (as in \cite{shchur2019intensity}), and mixture distributions can approximate any density on $\mathbb{R}$ arbitrarily well \cite{shchur2019intensity} not restricted to exponential or other monotonic functions. The conditional PDF is defined as,
\begin{equation}
p(\tau_i| w_i, \mu_i, s_i) = \sum_{k=1}^{K} w^k_i \frac{1}{\tau s^k_i \sqrt{2\pi}} \exp \left( -\frac{(\log \tau_i - \mu^k_i)^2}{2 (s^k_i)^2} \right)
\end{equation}
\begin{equation}
    w_i = \sigma(V_w c_i + b_w), \hspace*{0.2cm} s_i = \exp(V_s c_i + b_s), \hspace*{0.2cm} \mu_i = V_{\mu}  c_i + b_{\mu}
\end{equation}
where the mixture weights $w_i$, means $\mu_i$ and stddevs $s_i$ are parameterized by extracted context history $c_i$ and learnable $V, b$. 

The encoder-decoder parameters (denoted jointly as $\theta_a$) and the learnable account embeddings ($E$) can be learned using maximum likelihood estimation (for log-likelihood defined in Eq.~\ref{eq:amdn_likelihood}) from observed activity sequences and account set, and trained with gradient back-propagation as $\theta_a^*, E^* = \textrm{argmax}_{\theta_a,E}  \log p(C_s|U;\theta_a,E)$.  

\subsection{Modeling Hidden Groups} 
For modeling the hidden social groups from the observed activity traces, we model the hidden groups of accounts as a mixture of $N$ Gaussian multivariate distributions (GMM) in the account embedding space. 

Formally, the $i$-th social group is modeled as a Gaussian distribution $\mathcal{N}(\mu_i,\Sigma_i)$ where $\mu_i$ is the cluster mean and $\Sigma_i$ is the covariance matrix. Since the group membership of accounts is unknown, we assume that the account embeddings are drawn from the mixture, as embedding $E_{u_j}$ of account $u_j$ is distributed as $\sum_i p(i) \mathcal{N}(E_{u_j}; \mu_i, \Sigma_i)$, where $p(i)$ represents prior probability of group $i$. 

\textbf{Hidden group estimation:} A notable difference from general Gaussian mixture models, is that we define the GMM over the learnable account embeddings (\emph{latent} space), as compared to over \emph{observed} variables. Therefore, the optimization and learning of AMDN-HAGE requires bilevel optimization for jointly learning the model parameters and account embeddings (as discussed in the next section). The model proposed here aims to capture latent or hidden groups from activity traces, rather than from observed account features. This is because coordination patterns (like, activities of coordinated accounts might strongly influence each other) remain in activity traces and provide information about their collective behaviours, enabling identification of coordinated hidden groups.



For modeling distinct groups of coordinated and normal accounts, we used tied covariance $\Sigma$ for all groups. 
Denoting the parameters (means, the shared covariance, and prior) of GMM as $\theta_g$, the log-likelihood of social group modeling over accounts $U$ is:

\begin{align}
\begin{split}
\sum_{j=1}^{|U|}\log p(u_j;\theta_g,E) &= \sum_{j=1}^{|U|}\log\sum_{i=1}^{N}p(u_j,i;\theta_g, E)  \quad \\ 
& =\sum_{j=1}^{|U|}\log\sum_{i=1}^{N}p(i)\mathcal{N}(E_{u_j}; \mu_i, \Sigma)
\end{split}
\end{align}

\subsection{Jointly Learning}
For joint learning, we maximize following log-likelihood: 
\begin{equation}
\label{eq:amdn-hage-ll}
\begin{aligned}
    &\log p(C_s, U;\theta_g,\theta_a,E)= \log p(C_s| U;\theta_a,E) + \log p(U;\theta_g,E)\\
    &= \sum_{i=1}^{L} \left[ \log p_{\theta_a,E}(t_i|H_{t_i}) + \log p_{\theta_a,E}(u_i|H_{t_i}) \right]
    + \sum_{j=1}^{|U|}\log p(u_j;\theta_g,E)
\end{aligned}
\end{equation}
Above loss function has a trivial infinite asymptotic solution where all embedding vectors equal to the mean of a cluster and $\det(\Sigma)=0$. To avoid this solution, we constraint $\det(\Sigma)$ to be greater than a small constant $\lambda$, lower bounding the loss function. 

The log-likelihood (Eq~\ref{eq:amdn-hage-ll}) is a function of parameters $\theta_a, \theta_g, E$. In this, the optimization of the second term (Gaussian mixture of the shared latent embeddings), is a constrained optimization problem. Therefore, directly optimizing the joint likelihood with Stochastic Gradient Descent (SGD) or its variants like ADAM does not respect the constraints on mixture weights (normalized non-negative) and covariance (positive definite), leading to invalid log-likelihood in training (ablation study of loss with ADAM is in the expt section).

To address above disadvantages, we provide an equivalent bilevel optimization formulation to solve the joint learning problem:
\begin{equation}
    \theta_a^*, E^* = \textrm{argmax}_{\theta_a,E} [\log p(C_s|U;\theta_a,E) + \textrm{max}_{\theta_g}(\log p(U;\theta_g,E))]
\end{equation}
\begin{equation}
    \theta_g^* = \textrm{argmax}_{\theta_g}\log p(U;\theta_g,E^*) 
\end{equation}
Above bilevel optimization can be solved with iterative optimization. 
In each iteration, we first freeze $E$ and $\theta$, then estimate $\theta_g$ with EM algorithm. After that, we freeze $\theta_g$ and use SGD (or its variant) to optimize $E$ and $\theta_a$. Since the log-likelihood of \emph{latent} embeddings is optimized in the second term (Eq. \ref{eq:amdn-hage-ll}), we require initialization  of embeddings by pre-training $E$ and $\theta_a$ on observed sequences by maximizing the first term in the objective function, before jointly optimizing the two terms.
Detailed algorithm is in Algorithm \ref{alg:training}.

\begin{algorithm}[t]
  \caption{Training Algorithm for AMDN-HAGE}
  \label{alg:training}  
  \begin{algorithmic}[1]
    \REQUIRE Activity traces ($C_s$), Account set ($U$)
    \ENSURE Generative model ($\theta_a$, $\theta_g$ and $E$)
    \STATE $\theta_{a}^{(0)}, E^{(0)} \leftarrow \textrm{argmax}_{\theta_{a}, E} \log p(C_s|U;\theta_{a}, E)$
    \STATE Set $i$ as $1$ \COMMENT{Iteration index}.
    \WHILE{not converged}
    \STATE $\theta_{g}^{(i)} \leftarrow \textrm{argmax}_{\theta_{g}} \log p(U;E^{(i-1)},\theta_{g})$ using EM algorithm
    \STATE $\theta_{a}^{(i)}, E^{(i)} \leftarrow \textrm{argmax}_{\theta_a,E} \log p(C_s, U;\theta_g^{(i)},\theta_a,E)$ using SGD or its variants
    \STATE $i \leftarrow i+1$.
    \ENDWHILE
  \end{algorithmic}  
\end{algorithm} 

A generic concern to such alternating optimization algorithm is its convergence. Therefore, we give following theoretic guarantee that the proposed algorithm leads the model to converge at least on a local minimum or a saddle point with appropriate selection of the gradient based optimizer (denoting $L(\theta_a, E, \theta_g)$ as the negative log-likelihood (loss function)).
\begin{theorem}
Our proposed optimizing algorithm will converge at a local minimum or a saddle point if in any iteration $i$ the neural network optimizer satisfies following conditions:
\begin{itemize}
    \item Given the frozen $\theta_{g}^{(i)}$ acquired by EM algorithm in iteration $i$, the neural network optimization algorithm we applied in converges at a local minimum or or a saddle point $(\theta_{a}^{(i)},E^{(i)})$
    \item  $L(\theta_{a}^{(i)},E^{(i)},\theta_{g}^{(i)}) \leq L(\theta_{a}^{(i-1)},E^{(i-1)},\theta_{g}^{(i)})$, where
    $\theta_{a}^{(i-1)}$ and $E^{(i-1)}$ are the starting points in iteration $i$
\end{itemize}
\label{theorem:1}
\end{theorem}
The proof can be found in the Appendix. Theoretically, the two conditions can be guaranteed when our loss function is L-smooth and we apply standard Gradient Descent Algorithm with learning rate lower than $\frac{1}{L}$\cite{nesterov1998introductory}. But in practice, since finding strict local minimum is not as important as training speed and generalization, we can alternatively apply Adam or other variants.

%% file: expts.tex
\section{Experiment Results}

We verified the effectiveness of the proposed method AMDN-HAGE and training algorithm on real datasets collected from Twitter related to coordinated accounts by Russia’s Internet Research Agency, and for identification and analysis of coordination in COVID-19. We provide data collection, baselines and model variants. Details of implementation and experiment settings, are in the Appendix.\footnote{\href{ https://github.com/USC-Melady/
AMDN-HAGE-KDD21}{Our code is available here: \url{https://github.com/USC-Melady/AMDN-HAGE-KDD21}}}

\subsection{Data Collection}

\subsubsection{Russia's Internet Research Agency (IRA) coordinated campaign} 2752 Twitter accounts were identified by the U.S. Congress\footnote{https://www.recode.net/2017/11/2/16598312/russia-twittertrump-twitter-deactivated-handle-list} as coordinated accounts operated by the Russian agency (referred to as ``troll farm") to manipulate the U.S. Election in 2016. The social media posts (activities) of subset of these accounts were available through paid Twitter API access by the academic community to study coordinated account detection \cite{badawy2019characterizing,luceri2020detecting}. We obtained the collected dataset from \citeauthor{luceri2020detecting} which contains \textbf{312} of the Russian coordinated accounts (referred to as \emph{coordinated} \textbf{``trolls"}) with their 1.2M tweets, and also includes \textbf{1713 normal accounts} that participated in discussion about the U.S. Election during which the coordinated trolls were active (collected based on election related keywords using Twitter API); accounts in the collected dataset were active accounts with at least 20 active and passive tweets \footnote{Active tweet is where an account posts (tweet, retweet, reply) and passive is where the account is mentioned, retweeted, or replied \cite{luceri2020detecting}} \cite{luceri2020detecting}.

\textbf{Activity traces:} Activity traces $C_s$ are constructed from the tweets, as any account's posts and subsequent engagements from others (retweets, replies to the post) forming a time-ordered sequence of activities. The account set (2025 accounts) and activity traces are utilized to train AMDN-HAGE, and a held-out 15\% of sequence subsets are used as validation loss of the model.

\subsubsection{COVID-19 pandemic} Due to concerns around disinformation and social media abuse around COVID-19, we collect social media posts from with keywords related to COVID-19, using Twitter's streaming API service from March 1 to July 22, 2021. We use the collected \textbf{119,298 active accounts}, that have at least twenty active and passive collected tweets, and their \textbf{13.9M tweets}.

\textbf{Unknown coordination:} The COVID-19 data does \emph{not} have any labeled coordinated groups, unlike the IRA dataset. But it is important to examine if we can uncover any unknown coordinated campaigns. We run AMDN-HAGE on the full 119k account set with their activity traces as in the IRA dataset, and examine tweets from identified coordinated groups, and the overlap of accounts with suspended Twitter accounts (manually suspended by Twitter for violations of platform policies). There can be violating accounts that have not yet been found and suspended by Twitter. Also, accounts that are suspended can be due to various reasons (e.g. spam, automation, multiple accounts) but Twitter suspensions are not restricted to coordinating accounts. Thus we cannot use Twitter suspensions to estimate precision-recall on coordination detection.

\begin{table*}[t]
    \centering
    \caption{Results on detection of Russian coordinated disinformation campaign (IRA dataset) on Twitter in 2016 U.S. Election}
    \label{tab:IRA}
    \begin{tabular}{l|c|c|c|c|c|c|c}
    \toprule
       Method (Unsupervised) & AP & AUC & F1@TH=0.5 & Prec@TH=0.5 & Rec@TH=0.5 & MaxF1& MacroF1@TH=0.5 \\
    \midrule
     Co-activity &
    $0.208\pm0.01$&$0.592\pm0.03$&$0.292\pm0.02$&$0.206\pm0.02$&$0.510\pm0.04$&$0.331\pm0.03$&$0.515\pm0.02$ \\
    Clickstream & $0.169\pm0.02$&$0.535\pm0.04$&$0.215\pm0.06$&$0.205\pm0.05$&$0.228\pm0.08$&$0.215\pm0.06$&$0.532\pm0.03$ \\
    IRL & $0.200\pm0.00$&$0.610\pm0.02$&$0.265\pm0.02$&$0.219\pm0.02$&$0.336\pm0.03$&$0.340\pm0.02$&$0.543\pm0.01$ \\
    HP & $0.337\pm0.04$&$0.694\pm0.05$&$0.376\pm0.05$&$0.387\pm0.06$&$0.365\pm0.05$&$0.545\pm0.03$&$0.633\pm0.03$ \\
    \midrule
    AMDN + GMM &$0.787\pm0.05$&$0.894\pm0.03$&$0.631\pm0.06$&$\pmb{0.965\pm0.03}$&$0.472\pm0.07$&$0.738\pm0.05$&$0.792\pm0.03$ \\
    AMDN + Kmeans &$0.731\pm0.08$&$0.901\pm0.02$&$0.727\pm0.06$&$0.806\pm0.07$&$\pmb{0.663\pm0.06}$&$0.752\pm0.05$&$0.841\pm0.03$ \\
    \textbf{AMDN-HAGE} & $0.804\pm0.03$&$0.898\pm0.02$&$0.699\pm0.05$&$0.941\pm0.04$&$0.558\pm0.06$&$0.758\pm0.04$&$0.828\pm0.03$ \\
    \textbf{AMDN-HAGE} + Kmeans &$\pmb{0.818\pm0.04}$&$\pmb{0.935\pm0.02}$&$\pmb{0.731\pm0.04}$&$0.913\pm0.03$&$0.611\pm0.05$&$\pmb{0.776\pm0.03}$&$\pmb{0.846\pm0.02}$ \\
    \bottomrule
    Method (Supervised) & AP & AUC & F1@TH=0.5 & Prec@TH=0.5 & Rec@TH=0.5 & MaxF1& MacroF1@TH=0.5 \\
    \midrule
    IRL (S) & $0.672\pm0.08$&$0.896\pm0.03$&$0.557\pm0.06$&$\pmb{0.781\pm0.06}$&$0.436\pm0.06$&$0.633\pm0.07$&$0.749\pm0.03$ \\
    HP (S) & $0.760\pm0.04$&$0.925\pm0.02$&$0.753\pm0.02$&$0.743\pm0.04$&$0.769\pm0.06$&$0.782\pm0.03$&$0.853\pm0.01$ \\
    \midrule
    AMDN + NN &
    $0.814\pm0.04$&$0.918\pm0.02$&$0.733\pm0.04$&$0.710\pm0.05$&$0.761\pm0.05$&$0.763\pm0.04$&$0.841\pm0.02$ \\
    \textbf{AMDN-HAGE} + NN &
    $\pmb{0.838\pm0.04}$&$\pmb{0.926\pm0.03}$&$\pmb{0.769\pm0.04}$&$0.752\pm0.05$&$\pmb{0.789\pm0.05}$&$\pmb{0.799\pm0.04}$&$\pmb{0.862\pm0.02}$ \\
    \bottomrule 
    \end{tabular}
\end{table*}

\subsection{Baselines and Model Variant} We compare against existing approaches that utilize account activities to identify coordinated accounts. The baselines - extract features of coordination from account activities and use them for supervised or unsupervised detection of coordination, based on both individual or collective behaviours.
\begin{enumerate}
    \item Unsupervised Baselines: Co-activity clustering \cite{ruchansky2017csi} and Clickstream clustering \cite{wang2016unsupervised} are based on pre-defined activity features. Co-activity models joint activity in content sharing, and Clickstream clustering models patterns in post, retweet and reply actions. The SOTA approach is IRL \cite{luceri2020detecting} based on inverse reinforcement learning to extract features from activity traces, for clustering coordinated accounts. 
    \item Supervised Baselines: IRL(S) \cite{luceri2020detecting} is sota supervised variant of inverse reinforcement approach which trains a supervised classifier on extracted features from activity traces (with labeled subset of accounts).
\end{enumerate}

We add another baseline using HP (Hawkes Process) \cite{zhou2013learning} to learn account features unsupervised from activity traces. HP(S) is the supervised variant. HP models the influence between accounts with an additive function. This baseline serves as a ablation of the proposed model to show that latent influence and interaction patterns for coordinated group is more complex and neural point process can better extract these coordination features. 

For ablation of different components of proposed model AMDN-HAGE, we also  compare with (i) AMDN (without hiden group estimation), which only learns the activity trace model; we can use it to extract account embeddings and cluster with GMM or KMeans to identify the coordinated group as anomalous group. (ii) AMDN-HAGE directly uses the jointly learnt GMM to output the group membership. AMDN-HAGE + Kmeans instead uses account embeddings from AMDN-HAGE with KMeans clustering to find the anomalous coordinated group. (iii) To compare with supervised setting (IRL (S) \cite{luceri2020detecting}), we similarly train a classifier on extracted features i.e., learned account embeddings to detect coordinated accounts (assuming subset of labeled coordinated and normal accounts are available for training). The variants are AMDN + NN and AMDN-HAGE + NN which use a two-layer MLP classifier on extracted embeddings from AMDN and AMDN-HAGE resp.

\subsection{Results of Coordination Detection} 

\textbf{Detection results on IRA dataset:} We evaluate on two settings - unsupervised and supervised (as in earlier work \cite{luceri2020detecting}). In both, the the proposed model is trained as unsupervised from activity traces to obtain the group membership and account embeddings. 

In the unsupervised, the group membership is directly used to report anomalous coordinated group. In supervised, the learned embeddings are used as features to train a classifier (to predict coordinated from normal accounts). The classifier is trained on subset of labeled coordinated and normal accounts in IRA dataset, with rest (stratified 20\% in 5-folds) held-out for evaluation. 

\textbf{Table~\ref{tab:IRA}} provides results of model evaluation against the baselines averaged in 5-fold stratified cross-validation on the labeled normal and coordinated accounts in the IRA dataset over five random seeds. We compare the Average Precision (AP), area under the ROC curve (AUC), and F1, Precision, Recall, and MacroF1 at 0.5 threshold, and maxF1 at threshold that maximizes it. 
AMDN-HAGE
outperforms other methods on both unsupervised and supervised settings, due to its ability to capture coordination characteristics with diverse account behaviours by learning latent influence and hidden group behaviours, without pre-specifying features relied on by other baselines. 

Moreover, the coordination features learned with the proposed method are robust to unsupervised or supervised setting, unlike IRL and IRL(S) \cite{luceri2020detecting} (where even though IRL(S) can learn useful features, it performs poorly in unsupervised setting). In comparison, because AMDN-HAGE models the more \emph{intrinsic behaviours} of coordination, it can extract patterns that can effectively identify anomalous coordinated groups in an unsupervised manner. The margin is larger on unsupervised than supervised setting, where group behaviours are more important, since there is no known set of coordinated accounts to train classifiers from extracted features.




\textbf{Ablation of proposed model and training:} Besides baselines, we also compare AMDN-HAGE with its variants to verify the importance of the joint learning and optimization algorithm. 

To verify the importance of joint learning, in Table~\ref{tab:IRA}, AMDN-HAGE is compared with AMDN, which only learns the activity trace model, without hidden group estimation. Proposed model AMDN-HAGE captures consistently better coordination behaviours, indicating that modeling group or collective behaviour jointly is useful over only modeling the latent influence between account pairs through observed activity trace modeling.

\begin{figure}[t]
    \centering
  \begin{subfigure}{0.23\textwidth}
  \centering
\includegraphics[width=\textwidth]{ira_images/Adam-train.pdf}
    \caption{Validation loss with Adam leads to NaN loss function.}
\end{subfigure}
  \begin{subfigure}{0.23\textwidth}
  \centering
\includegraphics[width=\textwidth]{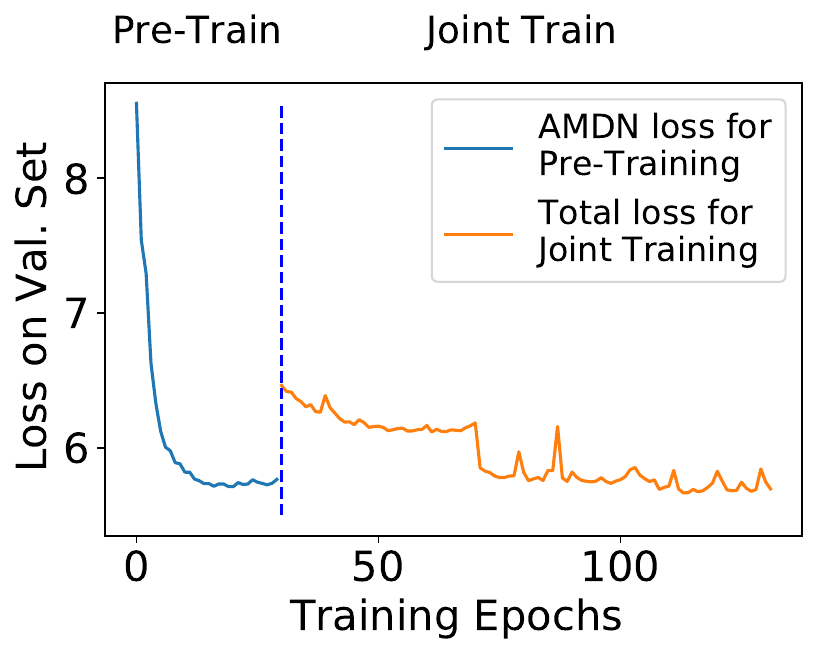}
    \caption{Loss on validation set in training proposed algorithm.}
\end{subfigure}
 \caption{Comparison of bilevel optimization and Adam.}
\label{fig:optimization}
\end{figure}

To demonstrate the effectiveness of the bilevel training algorithm, we present the validation loss (negative log-likelihood on held-out 15\% of activity traces) in the training process, comparing direct optimization of the joint log-likelihood using Adam (variant of SGD) and our proposed bilevel algorithm in Fig. \ref{fig:optimization}. As we can see, for the proposed optimization, the loss on the validation set in both the pre-training and joint training stage decline and finally converge. However, in direct optimization with Adam the validation loss decreases to a point but breaks as it reaches an invalid parameter point. Without constraints, Adam reaches a covariance matrix that is not positive definite, and an invalid log-likelihood (NaN).

\subsection{Analysis of Coordination Detection}
\begin{figure}[t]
    \centering
  \begin{subfigure}{0.2\textwidth}
  \centering
  \includegraphics[scale=0.25]{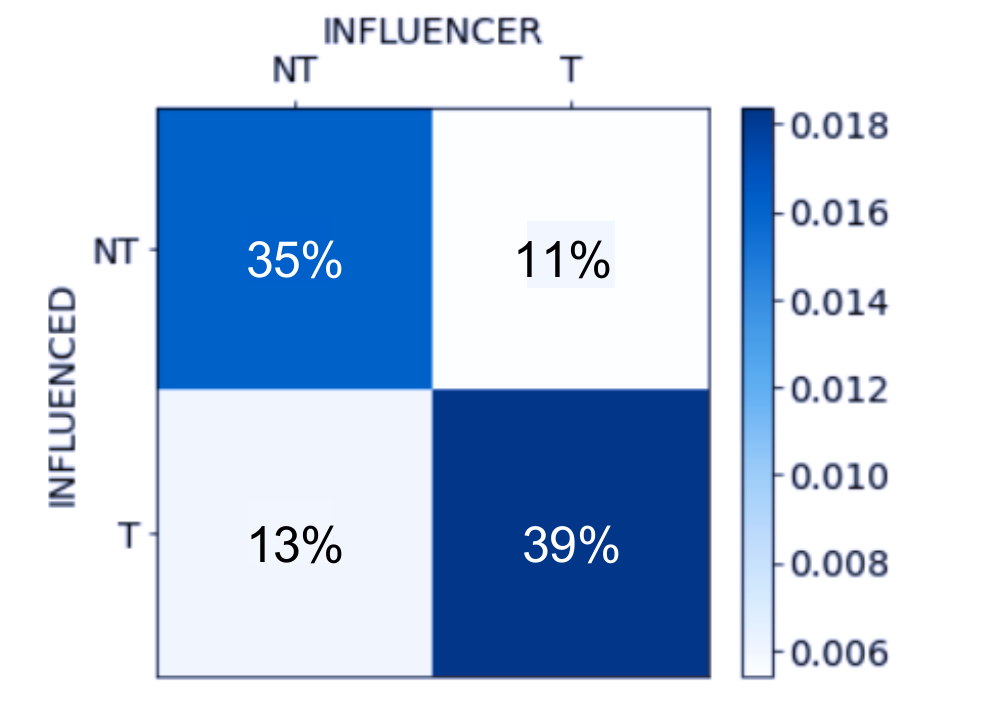}
    \caption{Avg. influence weights captured by AMDN-HAGE, which is highest for coordinated account pairs.}
    \label{fig:influence_IRA}
\end{subfigure}
 \hspace{0.02\textwidth}
  \begin{subfigure}{0.23\textwidth}
  \centering
  \includegraphics[scale=0.45]{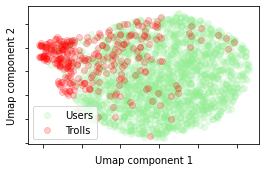}
    \caption{AMDN-HAGE account embeddings inferred for coordinated ``trolls" (red points) and normal accounts (green points).}
    \label{fig:embeddings_IRA}
\end{subfigure}
    \caption{Analysis of learned influence strength between Coordinated (``trolls" T) and Normal (non-trolls NT) in IRA.}
    \label{fig:ira_influence_emb_analysis}
\end{figure}
 \begin{figure}
     \begin{subfigure}{0.2\textwidth}
  \includegraphics[scale=0.27]{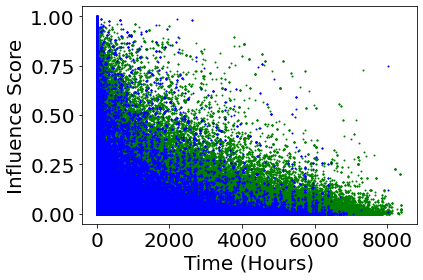}
    \caption{Overall trend on all account pairs.}
    \label{fig:nt-t-overall}
\end{subfigure}
 \hspace{0.02\textwidth}
\begin{subfigure}{0.2\textwidth}
  \centering
  \includegraphics[scale=0.27]{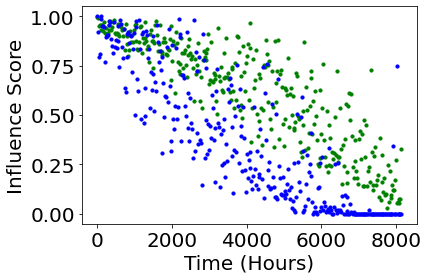}
    \caption{Trend of strongly interacting account pairs.}
    \label{fig:nt-t-max}
\end{subfigure}
     \caption{Analysis on how influence weights of account pairs vary with time difference. Green points: Normal (NT) account pairs. Blue: Coordinated (``trolls" T) account pairs.}
     \label{fig:nt-t}
 \end{figure}
 
\subsubsection{Uncovering characteristic behaviours from influence structure}

In this section, we examine the latent influence structure and account interactions learned by AMDN-HAGE on the IRA data. The latent influence is captured by the interpretable attention weights of the model, between account activities in observed traces. Higher attention paid by an event (account activity) on a history event (earlier activity from any account), indicates that the history event has a stronger triggering influence on the future event in consideration.

In Fig\ref{fig:influence_IRA}, we compute the aggregate influence between account pairs learned with AMDN-HAGE over 5 random seeds (as average attention weight from account interactions over all activity traces). The strongest influence ties are between coordinated (``trolls") (T) accounts, and the least influence is between normal (``non-trolls (NT)") and coordinated (T) accounts and their activities. In Figure~\ref{fig:embeddings_IRA}, the learned account embeddings of coordinated and normal accounts with AMDN-HAGE in the IRA data is visualized. As we see, coordinated accounts form an anomalous cluster distinct from normal accounts, reflecting our model captures coordinated behaviours.

In Fig.~\ref{fig:nt-t-overall} and \ref{fig:nt-t-max}, we examine how influence weights of account pairs vary with time difference. In the two figures, each point represent an account pair appearing in the same activity sequence. Blue points refer to coordinated (T) pairs and green points refer to normal account (NT) pairs. The influence weights (y-axis) correspond to the time difference between activities of two accounts appearing in the sequence (x-axis) is shown for all points in Fig.~\ref{fig:nt-t-overall} to reflect the overall trend. In Fig. \ref{fig:nt-t-max}, we only plot points with highest influence weight in each time difference range of 24 hrs to reflect the trend on strongly interacting pairs in each window. From both figures, we can see that influence weights are higher for shorter time differences between account activities. However, the influence decreases faster for coordinated pairs than normal account pairs with the passage of time between their activities. 

This phenomenon suggests that compared to normal social influence, the coordination is more temporal. Meanwhile, it is noticeable that a small fraction of coordination last even longer than 2000 hours (nearly 3 months), reflecting that the assumption on synchronization may fail in detecting such long-term coordination.

%% file: expts_covid.tex

\begin{table}[t]
    \centering
    \caption{Overlap between suspended Twitter accounts, and identified coordinated groups/ overall accounts in collected COVID-19 data.}
    \label{tab:twitter_labels_covid}
    \resizebox{.4\textwidth}{!}{
    \begin{tabular}{c|c|c|c}
    \toprule
    \textbf{Twitter} & \textbf{Overall} & \textbf{Cluster 1} & \textbf{Cluster 2} \\
   \textbf{Suspensions}& Overlap & Anomaly (3.7k) & Anomaly (5.5k) \\
  \midrule
    Suspended (9k) & 7.544 \% & 12.19 \% & 11.94 \% \\ \midrule
    State-backed (81) & 0.067 \% & 0.13 \% & 0.09 \% \\ \midrule
    Coupled (602) & 0.504 \% & 0.72 \% & 1.33 \% \\
    \midrule
    Targets (18.5k) & 15.507 \% & 14.98 \% & 17.11 \% \\
    \bottomrule
    \end{tabular}
    }
\end{table}

\begin{table}[t]
    \centering\small
 \caption{Representative tweets in disinformation topic clusters in identified COVID-19 coordinated accounts groups.}
    \label{tab:rep_tweets_covid}
    \renewcommand{\arraystretch}{0.5}
    \begin{tabular}{p{8.7cm}}
    \toprule
Did coronavirus leak from a research lab in Wuhan? Startling new theory is 'no longer being discounted' amid claims staff 'got infected after being sprayed with blood'  \#WWG1WGA \#QAnon \#MAGA \#Trump2020 \#COVID19  \\ 
\midrule
\#scamdemic Since China owns WHO, China must be in on the scam  The WHO Lied and Created a Global Panic: Second Extensive Study Finds Coronavirus Mortality Rate Is 0.4\% Not 3.4\% - Similar to Seasonal Flu  
\\ \midrule
VIRUS FRAUD - FAKE NEWS  In Spite of Leftist Media Hysteria at the Time, SD Governor Noem Confirms there were ZERO New Cases or 'Outbreaks' over Trump's Rushmore Event, Trump Supporters are Clean, Healthy People... 
\\ \midrule
BREAKING: \#BillGates Foundation And The \#Covid19 VACCINE NETWORK SCANDAL The British People Are Going To Get Very Angry Very Soon And Will Want ANSWERS.
\#GlaxoSmithKline \#BillGates  \#Rothschild \#WellcomeTrust \#COVID19  \#CivilService \#NoMasks
\\ \midrule
\#BillGates Negotiated \$100 Billion \#ContactTracing Deal With \#Democratic Congressman in 8/2019 well before \#Coronavirus \#Pandemic starts.  \#Gates holds pandemic drill 10/2019  Harvard finds \#SARSCOV2 started in 8/2019 in \#wuhan. Virus in USA by 1/2020. \\ 
\bottomrule
    \end{tabular}
\end{table}

\begin{figure}
    \centering
    \includegraphics[scale=0.38]{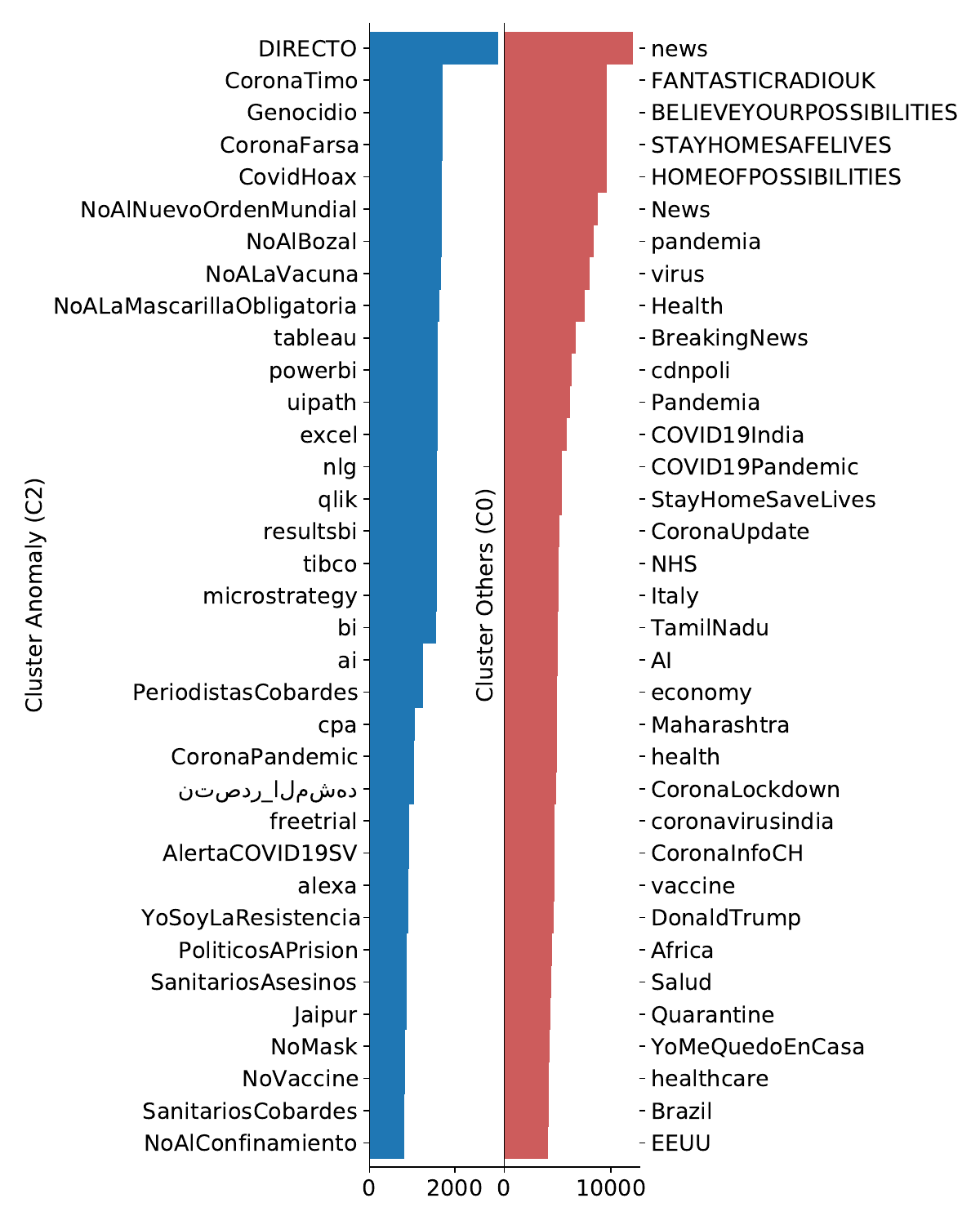}
    \caption{Top-35 (most frequent) unique hashtags in tweets of identified coordinated group and normal accounts.}
    \label{fig:dist_cluster_hashtags_main}
\end{figure}

\subsubsection{Uncovering coordinated groups in COVID-19 data} As mentioned earlier, the data collected on COVID-19 does not contain a ground-truth set of labeled coordinated accounts. But, we can use the proposed method AMDN-HAGE to uncover any suspicious coordinated accounts with analysis of features. AMDN-HAGE is trained on observed account activities in the COVID-19 data of $119k$ accounts. The method identifies two anomalous clusters of accounts (based on clustering silhouette scores, provided in appendix). We inspect the feature distribution in each account group.  

In Table~\ref{tab:twitter_labels_covid}, we compare the distribution of suspended Twitter accounts. Twitter additionally labels some suspended accounts as  state-backed i.e, accounts Twitter finds as linked to state-sponsored operations (such as from Russia, etc. that tried to interfere with politics in other countries) \cite{gadde2018enabling}. In addition, we consider the ``coupled" accounts in the collected data that bidirectionally engage with Twitter's state-backed accounts, and ``targets" as accounts mentioned (or targeted) by state-backed accounts in their tweets. Amongst all the 119k accounts, the distribution of Twitter accounts (Suspended, State-backed, Coupled) was between 1.5-2 times higher than by random chance in the identified anomalous clusters, even when the number of such accounts to be found from the large set of collected accounts is small. For Targets, the distribution is more uniform, since targeted accounts unlike Coupled only capture a unidirectional engagement attempt from state-backed accounts, which is expected because state-backed accounts attempt to manipulate and thus mention or engage with other normal accounts.

In Fig~\ref{fig:dist_cluster_hashtags_main}, we find most frequent hashtags in tweets posted by accounts in the groups, and plot the top hashtags unique to each group (hashtags of the smaller anomalous cluster is provided in the appendix). We find that the prominent top hundred hashtags in the coordinated group promote anti-mask and anti-vaccine (``NoMasks", ``NoVaccine", ``NoALaVacuna"), and anti-science theories (``Plandemic", ``Covid-Hoax"), and contain hashtags associated with ``QAnon" (``WWG1WGA"), a notorious far-right conspiracy group.

In Table~\ref{tab:rep_tweets_covid} we use topic modeling to find the most representative disinformation tweets posted by the anomalous accounts. We find 4 topic clusters within tweets that are linked to low-credibility (disinformation) news sources. The table shows tweets closest to the topic cluster centers. The narrative in the hashtags and topics suggests strong presence of tweets in Spanish and English (NoALaMascarillaObligatori, NoALaVacuna, NoAlNuevoOrdenMundia, NoVaccine) about no new world order, no masks, no vaccine, and QAnon, all opposing Bill Gates, and suggesting that COVID-19 is a hoax and deep state or political scam to monetize vaccines.  


%% file: conclusion.tex
\section{Conclusions}

In this work, we proposed a technique to detect coordinated accounts based on their collective behaviours, inferred directly from their activities on social media. The proposed method is independent of linguistic, metadata or platform specific features, and hence can be generalized across platforms and languages or countries from where disinformation campaigns originate. Such features can also be easily incorporated in the proposed model.

With analysis on Russian Interference and COVID-19 datasets, we investigated the behaviours of identified coordinated accounts, finding that influence between coordinated accounts is higher, and influence between coordinated pair of accounts decreases faster than non-coordinated pairs over time. In the COVID-19 data, we identified coordinated groups, and analysis suggests that the main narrative in the coordinated group is that COVID-19 is a hoax and political scam, with anti-vaccine, anti-mask social media posts.


%% file: suppl.tex
\section{Appendix}


\subsection{Details of the Baselines}
Here we provide the details of our baselines:
\begin{enumerate}
    \item Co-activity clustering \cite{ruchansky2017csi}. Co-activity features to identify coordinated groups as accounts that repeatedly share the same contents, was proposed in \citeauthor{ruchansky2017csi} Account features are extracted using SVD on a binary event-participation matrix (of accounts and its posts i.e., tweets, retweets, replies). Clustering on the features are used to detect coordination.
\item Clickstream clustering \cite{wang2016unsupervised,pacheco2020uncovering}. It is proposed by \citeauthor{wang2016unsupervised} to analyze account behaviours, and is based on hierarchical clustering of accounts with similar activity patterns to identify coordinated groups. Similarity between activities is represented based on post, reply and re-share patterns.
    \item IRL (S) \cite{luceri2020detecting}. SOTA approach is based on inverse reinforcement learning to discover motives of coordinated accounts from rewards estimated from activity traces. Estimated rewards are used as features for detection. IRL(S) trains AdaBoost on these features on a labeled subset of coordinated and normal accounts. \citeauthor{luceri2020detecting} reported different classifiers in their paper, including two-layer MLP and AdaBoost, with the latter outperforming others on their features.
    \item IRL \cite{luceri2020detecting}. We also compare an unsupervised variant of the SOTA IRL approach based on clustering with GMM or K-means clustering to detect coordinated groups.
    \item HP and HP (S). In addition, we compare our method to modeling activities using the Hawkes Process (HP)\footnote{Hawkes process code \cite{xu2018poppy} to extract account embeddings from activity traces} with influence $\alpha_{ij}$ factorized by learnable embeddings of accounts $i$ and $j$, with both clustering and supervised detection.
\end{enumerate}

\subsection{Details of Experiments}

\begin{figure}[h]
    \centering
    \includegraphics[width=.5\columnwidth]{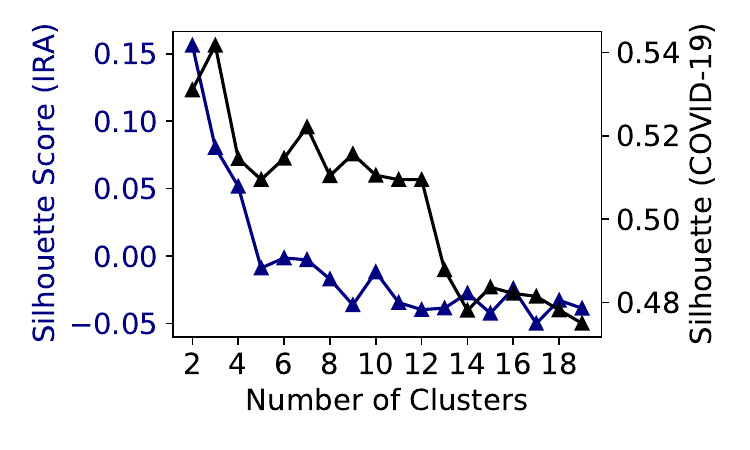}
    \caption{Selection of number of clusters based on silhouette scores in COVID-19 and IRA datasets.}
    \label{fig:sc}
\end{figure}

\textbf{Implementation details} We use activity sequences of maximum length 128, splitting longer sequences, batch size of 256 on 4 NVIDIA-2080Ti, embedding dimension in \{32, 64\}, number of mixture components for the PDF in the AMDN part between \{8,16,32\}, single head and single layer attention module. As for the component number in the HAGE part, it is set as 2 for IRA dataset and 3 for COVID dataset based on silhouette scores on learned embeddings from the pre-train step of training algorithm (Fig. \ref{fig:sc}). For the shared covariance matrix of HAGE part, we constrain it to be a diagonal matrix. The second term of the loss is scaled for embedding dimension and size of user set. We implement the model and training algorithm entirely in PyTorch and use Adam with 1e-3 learning rate and 1e-5 regularization to optimize. We train for max 1000 epochs with early stopping based on validation likelihood of sequences (75/15/10 splits). The link\footnote{\href{https://drive.google.com/drive/folders/1-vA2O76x6bxHNxPKjBGpndSdxOQhIZRT?usp=sharing}{Anonymized AMDN-HAGE code and COVID-19 data are available here (clickable).}} to anonymized code and collected COVID-19 data is in the footnote, and will be made non-anonymous if accepted. 

\begin{figure}
    \centering
    \includegraphics[width=.5\columnwidth]{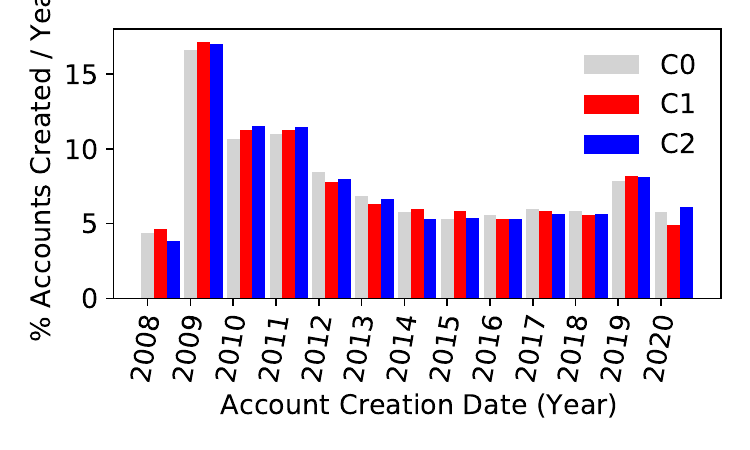}
    \caption{Distribution of account creation years in COVID-19 dataset for each identified accounts group or cluster.}
    \label{fig:covid_cluster_distri}
\end{figure}
\begin{figure*}
    \centering
    \includegraphics[width=.95\textwidth]{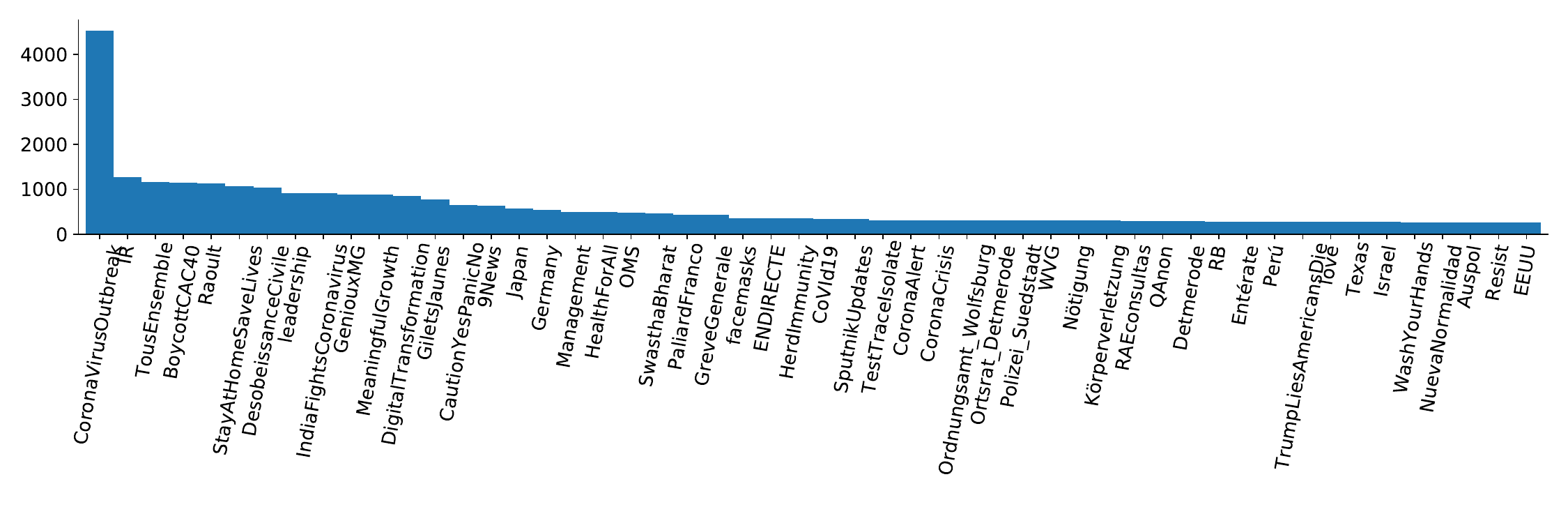}
    \caption{Hashtag distribution for cluster C1 (COVID-19 dataset).}
    \label{fig:hashtag}
\end{figure*}

Also, we present some numeric results and statistic properties we found on the COVID-19 dataset. 
We compare the distribution of account creation years in COVID-19 dataset for each identified accounts group or cluster. As we can see, the ratio of accounts created in recent years is higher in the two detected anomalous clusters, which is consistent to the reality that coordinated campaigns are raised in recent years. Also, we present the hashtag distribution in the C1 cluster in Figure \ref{fig:hashtag}. Surprisingly, although the ratio of suspended coordinated users in this cluster is significantly higher than normal, the hashtags look not like disinformation contents or topics. Upon further inspection, we find the political hashtags like ``Resist" or ``Resistance" in the top-50 unique hashtags of this cluster, refer to U.S. liberal political movement against the reigning president, and have not been linked to disinformation or conspiracies. This phenomenon suggests coordination may not be specific to disinformation spreading. In normal information spreading, there could be the influence from coordinated campaigns.

\subsection{Details of Neural Point Process Models} Methods SAHP \cite{zhangself} and THP \cite{zuo2020transformer} (Table ~\ref{tab:tpp} require Monte Carlo sampling to estimate an integral in the likelihood function used in MLE, as their intensity formulation does not have closed form solutions. This leads to more noisy approximation of gradients in training the model. On the other hand, HP \cite{zhou2013learning}, RMTPP \cite{du2016recurrent} define specific functional forms, e.g. exponential for intensity parameterization to have closed form likelihoods but are limited in the flexibility of the selected functional form of the intensity. LogNormMix \cite{shchur2019intensity} and FullyNN \cite{omi2019fully} alleviate these issues by modeling the conditional density and cumulative density getting flexibility and closed-form likelihoods. However, the neural network parameterizations in RMTPP, LogNormMix, FullyNN are based on recurrent neural networks and cannot be directly used to examine the influence between events and event types, which is possible in the other methods. We address this by modeling the conditional density function with interpretable neural network parameterizations, so that the influence structure can be learned and examined to understand coordinated behaviours of malicious accounts, and still have the same form for the point process conditional PDF as LogNormMix. 

\subsection{Proof of Convergence}

Here we provide the proof to Theorem \ref{theorem:1}.
\begin{proof}
We first prove that the training algorithm converges. In EM, each step increases the likelihood function of a mixture model. We have:
\begin{equation}
    \log P(E^{(i-1)}|\theta_{g}^{(i)}) \geq \log P(E^{(i-1)}|\theta_{g}^{(i-1)})
\end{equation}
Thus,  we obtain
\begin{equation}
    L(\theta_{a}^{(i-1)},E^{(i-1)},\theta_{g}^{(i)}) \leq L(\theta_{a}^{(i-1)},E^{(i-1)},\theta_{g}^{(i-1)}).
\end{equation}
Then, from the second condition, we know that:
\begin{equation}
    L(\theta_{a}^{(i)},E^{(i)},\theta_{g}^{(i)}) \leq L(\theta_{a}^{(i-1)},E^{(i-1)},\theta_{g}^{(i)})
\end{equation}
Therefore, we have:
\begin{equation}
    L(\theta_{a}^{(i)},E^{(i)},\theta_{g}^{(i)}) \leq L(\theta_{a}^{(i-1)},E^{(i-1)},\theta_{g}^{(i-1)})
\end{equation}
which means the loss function monotonically decreases. Since we constraint the variance of the mixture model in both point processing model and social group model larger than a constant $\epsilon$, the loss function is bounded by a constant $C$ on a given activity trace set:
\begin{equation}
    L(\theta_{a}^{(i)},E^{(i)},\theta_{g}^{(i)}) \geq C
\end{equation}
Thus the loss function converges when $i$ increases. Then we prove that the loss function converges at a local minimum or a saddle point. First when the parameters converges, we have:
\begin{equation}
    L(\theta_{a}^{(i)},E^{(i)},\theta_{g}^{(i)}) = L(\theta_{a}^{(i+1)},E^{(i+1)},\theta_{g}^{(i+1)}).
\end{equation}
Thus:
\begin{equation}
    L(\theta_{a}^{(i+1)},E^{(i+1)},\theta_{g}^{(i)}) = L(\theta_{a}^{(i+1)},E^{(i+1)},\theta_{g}^{(i+1)})
\end{equation}
\begin{equation}
    \log P(E^{(i)}|\theta_{g}^{(i+1)}) = \log P(E^{(i)}|\theta_{g}^{(i)}).
\end{equation}
In EM, if $\log P(E^{(i)}|\theta_{g}^{(i+1)}) = \log P(E^{(i)}|\theta_{g}^{(i)})$ then $\theta_{g}^{(i)} = \theta_{g}^{(i+1)}$.
Since $(\theta_{a}^{(i)},E^{(i)})$ is a local minimum or a saddle point, we have:
\begin{equation}
    \frac{\partial L(\theta_{a}^{(i)},E^{(i)},\theta_{g}^{(i)})}{\partial\theta_{a}^{(i)}} = \frac{\partial L(\theta_{a}^{(i)},E^{(i)},\theta_{g}^{(i)})}{\partial E^{(i)}} = 0
\end{equation}

Since EM is known to converge to a local minimum, we have:
\begin{equation}
    \frac{\partial \log P(E^{(i)}|\theta_{g}^{(i+1)})}{\partial \theta_{g}^{(i+1)}} = \frac{\partial \log P(E^{(i)}|\theta_{g}^{(i)})}{\partial \theta_{g}^{(i)}} = 0
\end{equation}
Therefore, $(\theta_{a}^{(i)},E^{(i)},\theta_{g}^{(i)})$ is a local minimum or a saddle point.
\end{proof}




%% file: sample-sigconf.bbl

\begin{thebibliography}{33}


\ifx \showCODEN    \undefined \def \showCODEN     #1{\unskip}     \fi
\ifx \showDOI      \undefined \def \showDOI       #1{#1}\fi
\ifx \showISBNx    \undefined \def \showISBNx     #1{\unskip}     \fi
\ifx \showISBNxiii \undefined \def \showISBNxiii  #1{\unskip}     \fi
\ifx \showISSN     \undefined \def \showISSN      #1{\unskip}     \fi
\ifx \showLCCN     \undefined \def \showLCCN      #1{\unskip}     \fi
\ifx \shownote     \undefined \def \shownote      #1{#1}          \fi
\ifx \showarticletitle \undefined \def \showarticletitle #1{#1}   \fi
\ifx \showURL      \undefined \def \showURL       {\relax}        \fi
\providecommand\bibfield[2]{#2}
\providecommand\bibinfo[2]{#2}
\providecommand\natexlab[1]{#1}
\providecommand\showeprint[2][]{arXiv:#2}

\bibitem[\protect\citeauthoryear{Addawood, Badawy, Lerman, and
  Ferrara}{Addawood et~al\mbox{.}}{2019}]%
        {addawood2019linguistic}
\bibfield{author}{\bibinfo{person}{Aseel Addawood}, \bibinfo{person}{Adam
  Badawy}, \bibinfo{person}{Kristina Lerman}, {and} \bibinfo{person}{Emilio
  Ferrara}.} \bibinfo{year}{2019}\natexlab{}.
\newblock \showarticletitle{Linguistic cues to deception: Identifying political
  trolls on social media}. In \bibinfo{booktitle}{\emph{ICWSM}}.
\newblock


\bibitem[\protect\citeauthoryear{Badawy, Addawood, Lerman, and Ferrara}{Badawy
  et~al\mbox{.}}{2019}]%
        {badawy2019characterizing}
\bibfield{author}{\bibinfo{person}{Adam Badawy}, \bibinfo{person}{Aseel
  Addawood}, \bibinfo{person}{Kristina Lerman}, {and} \bibinfo{person}{Emilio
  Ferrara}.} \bibinfo{year}{2019}\natexlab{}.
\newblock \showarticletitle{Characterizing the 2016 Russian IRA influence
  campaign}.
\newblock \bibinfo{journal}{\emph{SNAM}} \bibinfo{volume}{9},
  \bibinfo{number}{1} (\bibinfo{year}{2019}), \bibinfo{pages}{31}.
\newblock


\bibitem[\protect\citeauthoryear{Baumes, Goldberg, Magdon-Ismail, and
  Al~Wallace}{Baumes et~al\mbox{.}}{2004}]%
        {baumes2004discovering}
\bibfield{author}{\bibinfo{person}{Jeff Baumes}, \bibinfo{person}{Mark
  Goldberg}, \bibinfo{person}{Malik Magdon-Ismail}, {and}
  \bibinfo{person}{William Al~Wallace}.} \bibinfo{year}{2004}\natexlab{}.
\newblock \showarticletitle{Discovering hidden groups in communication
  networks}. In \bibinfo{booktitle}{\emph{ICISI}}. Springer.
\newblock


\bibitem[\protect\citeauthoryear{Cao, Yang, Yu, and Palow}{Cao
  et~al\mbox{.}}{2014}]%
        {cao2014uncovering}
\bibfield{author}{\bibinfo{person}{Qiang Cao}, \bibinfo{person}{Xiaowei Yang},
  \bibinfo{person}{Jieqi Yu}, {and} \bibinfo{person}{Christopher Palow}.}
  \bibinfo{year}{2014}\natexlab{}.
\newblock \showarticletitle{Uncovering large groups of active malicious
  accounts in online social networks}. In \bibinfo{booktitle}{\emph{Proceedings
  of the 2014 ACM Conference on Computer and Communications Security}}.
  \bibinfo{pages}{477--488}.
\newblock


\bibitem[\protect\citeauthoryear{Daley and Vere-Jones}{Daley and
  Vere-Jones}{2007}]%
        {daley2007introduction}
\bibfield{author}{\bibinfo{person}{Daryl~J Daley} {and} \bibinfo{person}{David
  Vere-Jones}.} \bibinfo{year}{2007}\natexlab{}.
\newblock \bibinfo{booktitle}{\emph{An introduction to the theory of point
  processes: volume II: general theory and structure}}.
\newblock \bibinfo{publisher}{Springer}.
\newblock


\bibitem[\protect\citeauthoryear{Du, Dai, Trivedi, Upadhyay, Gomez-Rodriguez,
  and Song}{Du et~al\mbox{.}}{2016}]%
        {du2016recurrent}
\bibfield{author}{\bibinfo{person}{Nan Du}, \bibinfo{person}{Hanjun Dai},
  \bibinfo{person}{Rakshit Trivedi}, \bibinfo{person}{Utkarsh Upadhyay},
  \bibinfo{person}{Manuel Gomez-Rodriguez}, {and} \bibinfo{person}{Le Song}.}
  \bibinfo{year}{2016}\natexlab{}.
\newblock \showarticletitle{Recurrent marked temporal point processes:
  Embedding event history to vector}. In \bibinfo{booktitle}{\emph{ACM
  SIGKDD}}. \bibinfo{pages}{1555--1564}.
\newblock


\bibitem[\protect\citeauthoryear{Ferrara, Varol, Davis, Menczer, and
  Flammini}{Ferrara et~al\mbox{.}}{2016}]%
        {ferrara2016rise}
\bibfield{author}{\bibinfo{person}{Emilio Ferrara}, \bibinfo{person}{Onur
  Varol}, \bibinfo{person}{Clayton Davis}, \bibinfo{person}{Filippo Menczer},
  {and} \bibinfo{person}{Alessandro Flammini}.}
  \bibinfo{year}{2016}\natexlab{}.
\newblock \showarticletitle{The rise of social bots}.
\newblock \bibinfo{journal}{\emph{Commun. ACM}} \bibinfo{volume}{59},
  \bibinfo{number}{7} (\bibinfo{year}{2016}), \bibinfo{pages}{96--104}.
\newblock


\bibitem[\protect\citeauthoryear{Gadde and Roth}{Gadde and Roth}{2018}]%
        {gadde2018enabling}
\bibfield{author}{\bibinfo{person}{Vijaya Gadde} {and} \bibinfo{person}{Yoel
  Roth}.} \bibinfo{year}{2018}\natexlab{}.
\newblock \showarticletitle{Enabling further research of information operations
  on Twitter}.
\newblock \bibinfo{journal}{\emph{Twitter Blog}}  \bibinfo{volume}{17}
  (\bibinfo{year}{2018}).
\newblock


\bibitem[\protect\citeauthoryear{Gupta, Kumaraguru, and Chakraborty}{Gupta
  et~al\mbox{.}}{2019}]%
        {gupta2019malreg}
\bibfield{author}{\bibinfo{person}{Sonu Gupta}, \bibinfo{person}{Ponnurangam
  Kumaraguru}, {and} \bibinfo{person}{Tanmoy Chakraborty}.}
  \bibinfo{year}{2019}\natexlab{}.
\newblock \showarticletitle{Malreg: Detecting and analyzing malicious retweeter
  groups}. In \bibinfo{booktitle}{\emph{Proceedings of the India Joint
  International Conference on Data Science and Management of Data}}.
  \bibinfo{pages}{61--69}.
\newblock


\bibitem[\protect\citeauthoryear{Hawkins}{Hawkins}{1980}]%
        {hawkins1980identification}
\bibfield{author}{\bibinfo{person}{Douglas~M Hawkins}.}
  \bibinfo{year}{1980}\natexlab{}.
\newblock \bibinfo{booktitle}{\emph{Identification of outliers}}.
  Vol.~\bibinfo{volume}{11}.
\newblock \bibinfo{publisher}{Springer}.
\newblock


\bibitem[\protect\citeauthoryear{Im, Chandrasekharan, Sargent, Lighthammer,
  Denby, Bhargava, Hemphill, Jurgens, and Gilbert}{Im et~al\mbox{.}}{2020}]%
        {im2020still}
\bibfield{author}{\bibinfo{person}{Jane Im}, \bibinfo{person}{Eshwar
  Chandrasekharan}, \bibinfo{person}{Jackson Sargent}, \bibinfo{person}{Paige
  Lighthammer}, \bibinfo{person}{Taylor Denby}, \bibinfo{person}{Ankit
  Bhargava}, \bibinfo{person}{Libby Hemphill}, \bibinfo{person}{David Jurgens},
  {and} \bibinfo{person}{Eric Gilbert}.} \bibinfo{year}{2020}\natexlab{}.
\newblock \showarticletitle{Still out there: Modeling and identifying russian
  troll accounts on twitter}. In \bibinfo{booktitle}{\emph{ACM Web Science}}.
  \bibinfo{pages}{1--10}.
\newblock


\bibitem[\protect\citeauthoryear{Luceri, Giordano, and Ferrara}{Luceri
  et~al\mbox{.}}{2020}]%
        {luceri2020detecting}
\bibfield{author}{\bibinfo{person}{Luca Luceri}, \bibinfo{person}{Silvia
  Giordano}, {and} \bibinfo{person}{Emilio Ferrara}.}
  \bibinfo{year}{2020}\natexlab{}.
\newblock \showarticletitle{Detecting troll behavior via inverse reinforcement
  learning: A case study of Russian trolls in the 2016 US election}. In
  \bibinfo{booktitle}{\emph{ICWSM}}, Vol.~\bibinfo{volume}{14}.
  \bibinfo{pages}{417--427}.
\newblock


\bibitem[\protect\citeauthoryear{Martin, Shapiro, and Nedashkovskaya}{Martin
  et~al\mbox{.}}{2019}]%
        {martin2019recent}
\bibfield{author}{\bibinfo{person}{Diego~A Martin}, \bibinfo{person}{Jacob~N
  Shapiro}, {and} \bibinfo{person}{Michelle Nedashkovskaya}.}
  \bibinfo{year}{2019}\natexlab{}.
\newblock \showarticletitle{Recent trends in online foreign influence efforts}.
\newblock \bibinfo{journal}{\emph{Journal of Information Warfare}}
  (\bibinfo{year}{2019}).
\newblock


\bibitem[\protect\citeauthoryear{Mei and Eisner}{Mei and Eisner}{2017}]%
        {mei2017neural}
\bibfield{author}{\bibinfo{person}{Hongyuan Mei} {and} \bibinfo{person}{Jason~M
  Eisner}.} \bibinfo{year}{2017}\natexlab{}.
\newblock \showarticletitle{The neural hawkes process: A neurally
  self-modulating multivariate point process}. In
  \bibinfo{booktitle}{\emph{NIPS}}. \bibinfo{pages}{6754--6764}.
\newblock


\bibitem[\protect\citeauthoryear{Nesterov}{Nesterov}{1998}]%
        {nesterov1998introductory}
\bibfield{author}{\bibinfo{person}{Yurii Nesterov}.}
  \bibinfo{year}{1998}\natexlab{}.
\newblock \showarticletitle{Introductory lectures on convex programming volume
  i: Basic course}.
\newblock \bibinfo{journal}{\emph{Lecture notes}} \bibinfo{volume}{3},
  \bibinfo{number}{4} (\bibinfo{year}{1998}), \bibinfo{pages}{5}.
\newblock


\bibitem[\protect\citeauthoryear{Omi, Aihara, et~al\mbox{.}}{Omi
  et~al\mbox{.}}{2019}]%
        {omi2019fully}
\bibfield{author}{\bibinfo{person}{Takahiro Omi}, \bibinfo{person}{Kazuyuki
  Aihara}, {et~al\mbox{.}}} \bibinfo{year}{2019}\natexlab{}.
\newblock \showarticletitle{Fully neural network based model for general
  temporal point processes}. In \bibinfo{booktitle}{\emph{NIPS}}.
  \bibinfo{pages}{2122--2132}.
\newblock


\bibitem[\protect\citeauthoryear{Pacheco, Hui, Torres-Lugo, Truong, Flammini,
  and Menczer}{Pacheco et~al\mbox{.}}{2021}]%
        {pacheco2020uncovering}
\bibfield{author}{\bibinfo{person}{Diogo Pacheco}, \bibinfo{person}{Pik-Mai
  Hui}, \bibinfo{person}{Christopher Torres-Lugo}, \bibinfo{person}{Bao~Tran
  Truong}, \bibinfo{person}{Alessandro Flammini}, {and}
  \bibinfo{person}{Filippo Menczer}.} \bibinfo{year}{2021}\natexlab{}.
\newblock \showarticletitle{Uncovering Coordinated Networks on Social Media}.
\newblock \bibinfo{journal}{\emph{ICWSM}}.
\newblock


\bibitem[\protect\citeauthoryear{Ruchansky, Seo, and Liu}{Ruchansky
  et~al\mbox{.}}{2017}]%
        {ruchansky2017csi}
\bibfield{author}{\bibinfo{person}{Natali Ruchansky}, \bibinfo{person}{Sungyong
  Seo}, {and} \bibinfo{person}{Yan Liu}.} \bibinfo{year}{2017}\natexlab{}.
\newblock \showarticletitle{CSI: A Hybrid Deep Model for Fake News Detection}.
  In \bibinfo{booktitle}{\emph{CIKM}}. ACM, \bibinfo{pages}{797--806}.
\newblock


\bibitem[\protect\citeauthoryear{Sharma, He, Seo, and Liu}{Sharma
  et~al\mbox{.}}{2021}]%
        {icwsmNetInference}
\bibfield{author}{\bibinfo{person}{Karishma Sharma}, \bibinfo{person}{Xinran
  He}, \bibinfo{person}{Sungyong Seo}, {and} \bibinfo{person}{Yan Liu}.}
  \bibinfo{year}{2021}\natexlab{}.
\newblock \showarticletitle{Network Inference from a Mixture of Diffusion
  Models for Fake News Mitigation}. In \bibinfo{booktitle}{\emph{ICWSM}}.
\newblock


\bibitem[\protect\citeauthoryear{Sharma, Qian, Jiang, Ruchansky, Zhang, and
  Liu}{Sharma et~al\mbox{.}}{2019}]%
        {sharma2019combating}
\bibfield{author}{\bibinfo{person}{Karishma Sharma}, \bibinfo{person}{Feng
  Qian}, \bibinfo{person}{He Jiang}, \bibinfo{person}{Natali Ruchansky},
  \bibinfo{person}{Ming Zhang}, {and} \bibinfo{person}{Yan Liu}.}
  \bibinfo{year}{2019}\natexlab{}.
\newblock \showarticletitle{Combating Fake News: A Survey on Identification and
  Mitigation Techniques}.
\newblock \bibinfo{journal}{\emph{ACM TIST}} (\bibinfo{year}{2019}).
\newblock


\bibitem[\protect\citeauthoryear{Sharma, Seo, Meng, Rambhatla, and Liu}{Sharma
  et~al\mbox{.}}{2020}]%
        {sharma2020coronavirus}
\bibfield{author}{\bibinfo{person}{Karishma Sharma}, \bibinfo{person}{Sungyong
  Seo}, \bibinfo{person}{Chuizheng Meng}, \bibinfo{person}{Sirisha Rambhatla},
  {and} \bibinfo{person}{Yan Liu}.} \bibinfo{year}{2020}\natexlab{}.
\newblock \showarticletitle{Coronavirus on social media: Analyzing
  misinformation in Twitter conversations}.
\newblock \bibinfo{journal}{\emph{arXiv preprint arXiv:2003.12309}}
  (\bibinfo{year}{2020}).
\newblock


\bibitem[\protect\citeauthoryear{Shchur, Bilo{\v{s}}, and G{\"u}nnemann}{Shchur
  et~al\mbox{.}}{2020}]%
        {shchur2019intensity}
\bibfield{author}{\bibinfo{person}{Oleksandr Shchur}, \bibinfo{person}{Marin
  Bilo{\v{s}}}, {and} \bibinfo{person}{Stephan G{\"u}nnemann}.}
  \bibinfo{year}{2020}\natexlab{}.
\newblock \showarticletitle{Intensity-Free Learning of Temporal Point
  Processes}.
\newblock \bibinfo{journal}{\emph{ICLR}}.
\newblock


\bibitem[\protect\citeauthoryear{Vaswani, Shazeer, Parmar, Uszkoreit, Jones,
  Gomez, Kaiser, and Polosukhin}{Vaswani et~al\mbox{.}}{2017}]%
        {vaswani2017attention}
\bibfield{author}{\bibinfo{person}{Ashish Vaswani}, \bibinfo{person}{Noam
  Shazeer}, \bibinfo{person}{Niki Parmar}, \bibinfo{person}{Jakob Uszkoreit},
  \bibinfo{person}{Llion Jones}, \bibinfo{person}{Aidan~N Gomez},
  \bibinfo{person}{{\L}ukasz Kaiser}, {and} \bibinfo{person}{Illia
  Polosukhin}.} \bibinfo{year}{2017}\natexlab{}.
\newblock \showarticletitle{Attention is all you need}. In
  \bibinfo{booktitle}{\emph{NeurIPS}}. \bibinfo{pages}{5998--6008}.
\newblock


\bibitem[\protect\citeauthoryear{Wang, Zhang, Tang, Zheng, and Zhao}{Wang
  et~al\mbox{.}}{2016}]%
        {wang2016unsupervised}
\bibfield{author}{\bibinfo{person}{Gang Wang}, \bibinfo{person}{Xinyi Zhang},
  \bibinfo{person}{Shiliang Tang}, \bibinfo{person}{Haitao Zheng}, {and}
  \bibinfo{person}{Ben~Y Zhao}.} \bibinfo{year}{2016}\natexlab{}.
\newblock \showarticletitle{Unsupervised clickstream clustering for user
  behavior analysis}. In \bibinfo{booktitle}{\emph{Proceedings of the 2016 CHI
  Conference on Human Factors in Computing Systems}}.
  \bibinfo{pages}{225--236}.
\newblock


\bibitem[\protect\citeauthoryear{Woolley and Howard}{Woolley and
  Howard}{2018}]%
        {woolley2018computational}
\bibfield{author}{\bibinfo{person}{Samuel~C Woolley} {and}
  \bibinfo{person}{Philip~N Howard}.} \bibinfo{year}{2018}\natexlab{}.
\newblock \bibinfo{booktitle}{\emph{Computational propaganda: political
  parties, politicians, and political manipulation on social media}}.
\newblock \bibinfo{publisher}{Oxford University Press}.
\newblock


\bibitem[\protect\citeauthoryear{Xu, Ruan, Korpeoglu, Kumar, and Achan}{Xu
  et~al\mbox{.}}{2019}]%
        {xu2019self}
\bibfield{author}{\bibinfo{person}{Da Xu}, \bibinfo{person}{Chuanwei Ruan},
  \bibinfo{person}{Evren Korpeoglu}, \bibinfo{person}{Sushant Kumar}, {and}
  \bibinfo{person}{Kannan Achan}.} \bibinfo{year}{2019}\natexlab{}.
\newblock \showarticletitle{Self-attention with functional time representation
  learning}. In \bibinfo{booktitle}{\emph{NeurIPS}}.
\newblock


\bibitem[\protect\citeauthoryear{Xu}{Xu}{2018}]%
        {xu2018poppy}
\bibfield{author}{\bibinfo{person}{Hongteng Xu}.}
  \bibinfo{year}{2018}\natexlab{}.
\newblock \showarticletitle{PoPPy: A Point Process Toolbox Based on PyTorch}.
\newblock \bibinfo{journal}{\emph{arXiv preprint arXiv:1810.10122}}
  (\bibinfo{year}{2018}).
\newblock


\bibitem[\protect\citeauthoryear{Yu, He, and Liu}{Yu et~al\mbox{.}}{2015}]%
        {yu2015glad}
\bibfield{author}{\bibinfo{person}{Rose Yu}, \bibinfo{person}{Xinran He}, {and}
  \bibinfo{person}{Yan Liu}.} \bibinfo{year}{2015}\natexlab{}.
\newblock \showarticletitle{Glad: group anomaly detection in social media
  analysis}.
\newblock \bibinfo{journal}{\emph{ACM Transactions on Knowledge Discovery from
  Data}} \bibinfo{volume}{10}, \bibinfo{number}{2} (\bibinfo{year}{2015}).
\newblock


\bibitem[\protect\citeauthoryear{Zannettou, Caulfield, De~Cristofaro,
  Sirivianos, Stringhini, and Blackburn}{Zannettou et~al\mbox{.}}{2019a}]%
        {zannettou2019disinformation}
\bibfield{author}{\bibinfo{person}{Savvas Zannettou}, \bibinfo{person}{Tristan
  Caulfield}, \bibinfo{person}{Emiliano De~Cristofaro},
  \bibinfo{person}{Michael Sirivianos}, \bibinfo{person}{Gianluca Stringhini},
  {and} \bibinfo{person}{Jeremy Blackburn}.} \bibinfo{year}{2019}\natexlab{a}.
\newblock \showarticletitle{Disinformation warfare: Understanding
  state-sponsored trolls on Twitter and their influence on the web}. In
  \bibinfo{booktitle}{\emph{Companion proceedings of WebConf}}.
  \bibinfo{pages}{218--226}.
\newblock


\bibitem[\protect\citeauthoryear{Zannettou, Caulfield, Setzer, Sirivianos,
  Stringhini, and Blackburn}{Zannettou et~al\mbox{.}}{2019b}]%
        {zannettou2019let}
\bibfield{author}{\bibinfo{person}{Savvas Zannettou}, \bibinfo{person}{Tristan
  Caulfield}, \bibinfo{person}{William Setzer}, \bibinfo{person}{Michael
  Sirivianos}, \bibinfo{person}{Gianluca Stringhini}, {and}
  \bibinfo{person}{Jeremy Blackburn}.} \bibinfo{year}{2019}\natexlab{b}.
\newblock \showarticletitle{Who let the trolls out? towards understanding
  state-sponsored trolls}. In \bibinfo{booktitle}{\emph{ACM WebSci}}.
  \bibinfo{pages}{353--362}.
\newblock


\bibitem[\protect\citeauthoryear{Zhang, Lipani, Kirnap, and Yilmaz}{Zhang
  et~al\mbox{.}}{2020}]%
        {zhangself}
\bibfield{author}{\bibinfo{person}{Qiang Zhang}, \bibinfo{person}{Aldo Lipani},
  \bibinfo{person}{Omer Kirnap}, {and} \bibinfo{person}{Emine Yilmaz}.}
  \bibinfo{year}{2020}\natexlab{}.
\newblock \showarticletitle{Self-Attentive Hawkes Process}.
\newblock \bibinfo{journal}{\emph{ICML}}.
\newblock


\bibitem[\protect\citeauthoryear{Zhou, Zha, and Song}{Zhou
  et~al\mbox{.}}{2013}]%
        {zhou2013learning}
\bibfield{author}{\bibinfo{person}{Ke Zhou}, \bibinfo{person}{Hongyuan Zha},
  {and} \bibinfo{person}{Le Song}.} \bibinfo{year}{2013}\natexlab{}.
\newblock \showarticletitle{Learning social infectivity in sparse low-rank
  networks using multi-dimensional hawkes processes}. In
  \bibinfo{booktitle}{\emph{AISTATS}}.
\newblock


\bibitem[\protect\citeauthoryear{Zuo, Jiang, Li, Zhao, and Zha}{Zuo
  et~al\mbox{.}}{2020}]%
        {zuo2020transformer}
\bibfield{author}{\bibinfo{person}{Simiao Zuo}, \bibinfo{person}{Haoming
  Jiang}, \bibinfo{person}{Zichong Li}, \bibinfo{person}{Tuo Zhao}, {and}
  \bibinfo{person}{Hongyuan Zha}.} \bibinfo{year}{2020}\natexlab{}.
\newblock \showarticletitle{Transformer Hawkes Process}.
\newblock \bibinfo{journal}{\emph{NeurIPS}}.
\newblock


\end{thebibliography}
